\documentclass[numberedappendix]{emulateapj}

\usepackage{graphicx}
\usepackage{epstopdf}

\shorttitle{ Mg II -- LRG Cross-Correlation Analysis}
\shortauthors{Lundgren et al.}

\begin{document}


\title{A Cross-Correlation Analysis of  Mg II Absorption Line Systems and Luminous Red Galaxies from the SDSS DR5}


\author{
Britt F. Lundgren\altaffilmark{1}, Robert J. Brunner\altaffilmark{1,2},  Donald G. York\altaffilmark{3,4}, Ashley J. Ross\altaffilmark{1}, Jean M. Quashnock\altaffilmark{3,5}, Adam D. Myers\altaffilmark{1},  Donald P. Schneider\altaffilmark{6}, Yusra AlSayyad\altaffilmark{4}, Neta Bahcall\altaffilmark{7}
}
\altaffiltext{1}{Department of Astronomy, University of Illinois, Urbana, IL 61801}
\altaffiltext{2}{National Center for Supercomputing Applications, Champaign, IL 61820} 
\altaffiltext{3}{Department of Astronomy and Astrophysics, University of Chicago, Chicago, IL 60637}
\altaffiltext{4}{Enrico Fermi Institute, University of Chicago, Chicago, IL 60637} 
\altaffiltext{5}{Department of Physics, Carthage College, Kenosha, WI  53140}
\altaffiltext{6}{Department of Astronomy and Astrophysics, The Pennsylvania State University, 525 Davey Laboratory, University Park, PA 16802}
\altaffiltext{7}{Princeton University Observatory, Peyton Hall, Princeton, NJ 08544}

\begin{abstract}

We analyze the cross-correlation of  Mg II ($\lambda2796, 2803$) quasar absorption systems with luminous red galaxies (LRGs) from the Fifth Data Release of the Sloan Digital Sky Survey.  The absorption line sample consists of 2,705 unambiguously intervening   Mg II absorption systems, detected at a 4$\sigma$ level, covering a redshift range (0.36$\leq$z$_{abs}$$\leq$0.8) and a rest equivalent width range of 0.8\AA $ $ $\leq$ W$_{r}^{\lambda2796}\leq$5.0\AA.   We cross-correlate these absorbers with 1,495,604 LRGs with accurate photometric redshifts in the same redshift range and examine the relationship of  Mg II equivalent width and clustering amplitude.  We confirm with high precision a previously reported weak anti-correlation of equivalent width and the dark matter halo mass, measuring dark matter halo masses of Mg II absorbers to be log$M_{h}(M_{\sun}h^{-1})=11.29\pm^{0.36}_{0.62}$ for the W$_{r}$$\ge$1.4\AA $ $ sample and log$M_{h}(M_{\sun}h^{-1})=12.70\pm^{0.53}_{1.16}$  for absorbers with 0.8\AA$\leq W_{r}<$1.4\AA $ $. These measurements agree with previous reported values within the stated errors.  Additionally, we investigate the significance of a number of potential sources of bias inherent in absorber-LRG cross-correlation measurements, including absorber velocity distributions and weak lensing of background quasars, which we determine is capable of producing a 20--30\%  bias in angular cross-correlation measurements on scales less than 2$\arcmin$.  We measure the Mg II -- LRG cross-correlation for 719 absorption systems with $v<60,000$ km s$^{-1}$ in the quasar rest frame and find that these absorbers typically reside in dark matter haloes that are $\sim$10--100 times more massive than those hosting unambiguously intervening Mg II absorbers.  Furthermore, we find evidence for evolution of the redshift number density, $\partial N / \partial z$, with 2$\sigma$ significance for the strongest (W$_{r}^{\lambda2796}\gtrsim2.0$\AA) absorbers in the DR5 sample.  This width-dependent $\partial N / \partial z$ evolution does not significantly affect the recovered equivalent width--halo mass anti-correlation and adds to existing evidence that the strongest  Mg II absorption systems are correlated with an evolving population of field galaxies at these redshifts, while the non-evolving $\partial N / \partial z$ of the weakest absorbers more closely resembles the LRG population.

\end{abstract}


\section{Introduction}

Owing to the advent of large spectroscopic surveys such as the Sloan Digital Sky Survey (SDSS; York et al. 2000), tens of thousands of quasar absorption lines (QALs) have now been identified. The detection of QALs is largely independent of the luminosity of the quasars, and as a result QAL catalogs extend to $z\sim6$ and hold a unique potential for probing the large-scale structure of baryonic matter in the high-redshift Universe.  Despite the existence of new large data sets, many critical questions regarding the origins and environments of QALs remain unanswered.  QALs have a well-established association with $\sim L^{*}$ galaxies (Bergeron 1986; Lanzetta \& Bowen 1990, 1992; Steidel $\&$ Sargent 1992; Steidel et al. 1994; Zibetti et al. 2005; Nestor et al. 2007; Kacprzak et al. 2007); but whether QALs originate in cool extended regions of dark-matter haloes (as first proposed by Bahcall \& Spitzer 1969 and later developed by Mo \& Miralda-Escude 1996, Maller \& Bullock 2004, and Chelouche et al. 2008)  or as a result of super-winds from phenomena within galactic disks (see, e.g., Nestor et al. 2005; Bouch{\'e} et al. 2006; Zibetti et al. 2007) remains a matter of debate.  Determining the origins and environments of QALs at redshifts that overlap with observable galaxies is a critical first step toward using quasar absorption line catalogs to examine the content, structure, and evolution of gas in the high-redshift Universe.

The recent releases of large QAL catalogs derived from SDSS spectra (Prochaska \& Herbert-Fort 2004; Bouch{\'e} et al. 2007; York et al. 2009, in prep.) and an ever-improving understanding of the various classifications of QALs have facilitated the statistical examination of the environments of QALs.  Clustering measurements in particular have proven to be powerful probes of QAL environments, revealing their typical underlying dark matter halo masses (e.g., Bouch{\'e} et al 2004, 2005, 2006; Cooke et al. 2006; Ryan-Webber 2006; Wilman et al. 2007) and the fraction and distribution of absorbers intrinsic to quasar outflows (Nestor et al. 2008; Wild et al. 2008).  In a Universe dominated by cold dark matter, galaxies are found in dark matter haloes with a structure determined by universal scaling relations (e.g., Navarro, Frenk \& White 1996) and a clustering amplitude that scales with halo mass (Mo \& White 1996).  Since baryons trace this underlying distribution, their clustering is also determined by the mass of the dark matter haloes in which they reside.  As a result, measurements of the clustering of baryonic matter can be used to extract typical dark matter halo masses (Mo \& White 1996; Sheth, Mo, \& Tormen 2001).  

Due to the low space-densities of quasars, QALs are somewhat sparsely sampled. Thus, the QAL auto-correlation can be difficult to measure.  However, as demonstrated by Bouch{\'e} et al. (2004, 2006), by measuring the cross-correlation of QALs with better-sampled galaxy populations with well-known dark matter halo masses one can both significantly increase the signal-to-noise of clustering measurements and more simply determine the dark matter halo mass environments of QALs.

 Mg II ($\lambda$2796,2803) absorption is a common and easily identifiable feature in quasar spectra, which is detected in optical wavelengths over a redshift range $0.3<z<2.2$, making it the best QAL candidate for cross-correlations with luminous galaxies detected photometrically by the SDSS.   Mg II is known to trace cold photo-ionized gas with T$\sim10^{4}$K (Lanzetta \& Bowen 1990; Hamann 1997) and has been observed in association with five decades of HI column density, $10^{17} $cm$^{-2}\leq N_{HI}\leq10^{22}$cm$^{-2}$ (Bergeron \& Stasinska 1986; Steidel \& Sargent 1992; Churchill et al. 2000).  The detection of  Mg II over a range of galactic impact parameters extending to $\sim$100 h$^{-1}$ kpc  (Bergeron et al. 1987; Steidel et al. 1995; Churchill et al. 2005; Zibetti et al. 2007) also makes this species of QAL a powerful probe of a wide range of galactic environments.

Luminous red galaxies (LRGs) provide a well-suited galaxy sample for cross-correlations with  Mg II since they are abundant and have typical luminosities allowing for their observation in the SDSS to redshift z$\sim$0.8.  LRGs can be photometrically selected by well understood color criteria (e.g., Eisenstein et al. 2001; Cannon et al. 2006), and methods for determining the photometric redshifts of these galaxies are highly reliable, with an average rms photometric redshift accuracy of $\sigma_{z}=0.05$ (Padmanabhan et al. 2005; Collister et al. 2007).  LRGs have been shown to be strongly clustered (e.g.,  Eisenstein et al. 2005a; Zehavi et al. 2005; Ross et al. 2007; Blake et al. 2008; Wake et al. 2008), and they exhibit no significant evolution in their stellar mass and clustering amplitude over much of the observable redshift range in the SDSS (Wake et al. 2006; 2008a; Brown et al. 2007; Cool et al. 2008; Brown et al. 2008).  As a result, the typical dark matter halo masses of LRGs have also been calculated to high precision (Blake et al. 2008; Wake et al. 2008a; Zheng et al. 2008).

Bouch{\'e} et al. (2004) cross-correlated  Mg II and LRGs extracted from the SDSS First Data Release (DR1; Strauss et al. 2002; Abazajian et al. 2003) and found the average halo mass for $W_{r}>0.3$\AA $ $  Mg II QALs to be consistent with 0.7 L$_{*}$ galaxies.  Bouch{\'e} et al. (2006) further reported an anti-correlation between rest equivalent width of the 2796\AA $ $ line ($W_{r}^{\lambda2796}$) and the clustering amplitude of  Mg II absorbers with LRGs in the SDSS Third Data Release (DR3; Abazajian et al. 2005).  The latter result was surprising, since equivalent widths of QALs have been shown to increase in proportion to the number of components in the absorbing system (Petitjean $\&$ Bergeron 1990; Churchill $\&$ Vogt 2001; Churchill, Vogt, \& Charlton 2003; Prochter et al. 2006), and classical models predict that the width should also be directly related to the velocity dispersion, which increases with dark matter halo mass.  Bouch{\'e} et al. (2006), hereafter B06, interpret their result as evidence supporting a model where strong absorption lines are not produced by virialized gas in large extended galaxy haloes but instead originate in supernovae-driven winds.  This interpretation has been augmented by the recent results by Zibetti et al. (2007), which used SDSS data and image-stacking techniques to link strong  Mg II absorbers with blue star-forming galaxies and weak  Mg II absorbers with red passive galaxies.  In a related finding, Nestor et al. (2007) have provided specific examples of associations between the strongest Mg II absorbers (W$_{r}^{2796}$\AA$>2.7$\AA) and galaxies that have evidence of either recent interaction or starburst activity.

Tinker \& Chen (2008) suggest an alternate interpretation of the apparent halo mass---equivalent width anti-correlation, proposing that the higher temperatures of very massive halos limit the amount of available cold gas, which is required to produce strong Mg II QALs.  Within this model, the overall fraction of strong Mg II absorbers may be preferentially reduced in the most massive halo environments, thereby suppressing the overall clustering of the strongest absorbers.  Tinker \& Chen (2008) argue this effect may explain the halo mass -- equivalent-width anti-correlation without evoking feedback mechanisms from active star formation.  

The growing debate over the results and interpretation of the equivalent width---halo mass anti-correlation has provided strong motivation to verify the relationship between  Mg II equivalent width and halo mass.  As a result, we have measured the  Mg II -- LRG cross-correlation using the largest and most precise catalogs of photometric LRGs and intervening Mg II absorption systems to date, which we have independently derived from the SDSS DR5.  We have also examined the significance of a number of potential biases not explored in earlier analyses, including the effects of weak lensing and contamination of the Mg II sample by associated absorption systems.  For the remainder of this work, we assume a flat cosmology with $\Omega_{M}=0.3$, $\Omega_{\Lambda}=0.7$, and $h_{o}=1.0$. 
  
\section{Observations and Data Reduction}

\subsection{The Sloan Digital Sky Survey\label{SDSS}}

The data for this paper have been drawn from the Fifth Data Release (DR5) of the SDSS \citep{AM07}. Through June 2005, the SDSS had imaged $\gtrsim$8000 deg$^{2}$ and obtained follow-up spectra for nearly $7 \times 10^{5}$ galaxies and $8 \times 10^{4}$ quasars.  Imaging data are acquired by a drift-scan camera with 30 photometric chips and 24 astrometric chips \citep{gunn98, Gunn06} on the dedicated 2.5-meter telescope at Apache Point Observatory in New Mexico.  The data are reduced and calibrated by the PHOTO software pipeline \citep{lupton02}.   The photometric system is normalized such that the SDSS $u,g,r,i$ and $z$ magnitudes \citep{fukugita96} are on the AB system \citep{smith02}.  A 0.5-meter telescope monitors site photometric quality and extinction \citep{hogg01}.  Photometric calibration is achieved using the Monitor Telescope Pipeline, described in \citet{tucker06}. Point source astrometry for the survey is accurate to less than $100$ milliarcseconds \citep{pier03}, and imaging quality control is discussed in \citet{ivezic042}.

A fraction of the objects located in the imaging are targeted for spectroscopy as candidate galaxies \citep{eisenstein01, strauss02}, quasars \citep{richards02b}, or stars \citep{Stoughton02}.  Targeted objects are grouped in 3-degree diameter tiles \citep{Blanton03} and aluminum plates are drilled with 640 holes at positions corresponding to the location of the objects on the sky.  When the telescope is in spectroscopic mode, roughly 500 galaxies, 50 quasars and 50 stars are observed on each plate, and the remaining fibers are utilized for sky subtraction and calibration.

SDSS spectra cover the observer-frame optical and near infrared, from 3900\AA--9100\AA, with a resolution of $\lambda \over \Delta \lambda$ $\approx$ 2000 at 5000\AA\ \citep{Stoughton02}. Spectra are obtained in a series of consecutive 15-minute observations until an average minimum signal-to-noise ratio is met.  Observations of 32 sky fibers, 8 reddening standard stars, and 8 spectrophotometric standard stars are used to calibrate the spectra of the science targets.  The {\tt Spectro2d} pipeline flat-fields and flux calibrates the spectra and the {\tt Spectro1d} code identifies spectral features and classifies objects by spectral type \citep{Stoughton02}.  Ninety-four percent of all SDSS quasars are identified spectroscopically by this automated calibration; the remaining quasars are identified through visual inspection.  Quasars are defined to be those extragalactic objects with broad emissions lines (full width at half maximum velocity width $\gtrsim 1000$ km s,$^{-1}$ regardless of luminosity).  Objects meeting these criteria have been compiled in the SDSS DR5 quasar catalog \citep{S07}.  This catalog covers an area of approximately 5740 deg$^{2}$ and includes 77,429 quasars with $i$-band magnitudes fainter than 15.0 and highly-reliable redshifts in the range 0.08 $\leq$ z $\leq$ 5.41.  The SDSS Quasar catalog also includes a luminosity limit of M$_{i}$=-22.0 for a cosmology with $H_{0}$ = 70 km s$^{-1}$ Mpc$^{-1}$, $\Omega_{M}$ = 0.3, and $\Omega_{\Lambda}$ = 0.7.

\subsection{ Mg II Absorption Systems}

Most work published prior to the onset of large-scale surveys like the SDSS focused on the statistics of a small number of high resolution spectra, which could be examined by individual inspection.  However, a reliable and fully automated process was required to analyze the large number of quasar spectra obtained with the SDSS.  A number of algorithms exist to automatically identify absorption systems given a list of observed lines (see, e.g., Bahcall 1968; Schneider et al 1993), but these have generally been tuned for the analysis of spectra with relatively high resolution.  A more complex algorithm is required to identify QALs in SDSS spectra, which have lower average signal-to-noise and resolution than previous quasar absorption line data.  By developing an independent algorithm to detect systems within the SDSS spectra and quantify the reliability of each detection, York et al. 2009 (hereafter, Y09), have cataloged tens of thousands of absorption systems without the bias introduced with a by-eye examination.  The automated process used to identify and grade absorption systems is detailed extensively in Y09; for completeness, we present a brief summary of this process below.  

The Y09 quasar absorption line catalog detects QALs using two primary codes: an absorption line-finding algorithm and a code that sorts detected lines into systems of the same redshift.  The first code normalizes each quasar spectrum to an average continuum and records the equivalent widths for absorption features with $\frac{W_{r}}{\sigma_{W_{r}}} \geq3$.  A system-finding algorithm then operates on the entire list of $3\sigma$ detected lines, matching different ions with the same redshifts.  The first task of this algorithm is to identify easily observable ion doublets of C IV and Mg II with redshifts foreground to the quasar.  Since it is not uncommon to observe intrinsic absorption features that are slightly blueshifted with respect to the quasar emission, we also allow systems to be identified with blueshifted velocities as great as -6000 km s$^{-1}$ in the quasar rest frame.  Doublets are retained if the wavelength separation matches that expected for C IV ($\lambda$1548.2, 1550.8) or Mg II ($\lambda$2796.4, 2803.5), redshifted within the respective ranges observable in the SDSS: 1$.5<z_{C IV}<4.5$; $0.36<z_{ Mg II}<2.2$.  Redshifts are calculated separately for each line in a candidate doublet.  If the difference in these redshifts is smaller than the FWHM of the first line divided by its rest wavelength, the doublet is kept.  We further require that the line pair exhibits a reasonable doublet ratio, defined as:
\begin{equation}
DR = \frac{f_{2}\lambda_{r_{2}}^{2}}{f_{1}\lambda_{r_{1}}^{2}},   
\end{equation}
where $f_{1}$ and $f_{2}$ indicate the $f$-value for each line in the doublet with rest wavelengths $\lambda_{r}$ (Morton 2003), labeled in order of increasing wavelength.  For lines with $\frac{W}{\sigma_{W}}<20$, we require:
\begin{equation}
W_{2}-\sigma_{W_{2}}\leq W_{1}+\sigma_{W_{1}}, 
\end{equation}
and
\begin{equation}
\frac{W_{2}+\sigma_{W_{2}}}{W_{1}}\geq DR
\end{equation}
where $W_{1}$ and $W_{2}$ represent the equivalent widths of each line in the candidate doublet, again labeled in order of increasing wavelength.  In cases where both lines in a doublet are strongly saturated, with $\frac{W}{\sigma_{W}}\geq20$, we require:
\begin{equation}
\frac{W_{2}-3\sigma_{W_{2}}}{W_{1}}\geq DR
\end{equation}

The algorithm then assigns a redshift to the positively identified doublet and sweeps through the remaining list of detected lines to identify other ions at the same redshift.  Lines matching to a system within an allowed redshift offset:
\begin{equation}
dz\leq\mid\frac{FWHM}{\lambda_{r}}\mid
\end{equation}
are then grouped into absorption systems and cataloged.  For the purpose of defining complete samples, doublets of C IV and  Mg II are searched for independently, although they are often detected in the same systems.  

All absorption systems in the Y09 catalog have been assigned grades, which provide a measurement of their reliability based on the number of different ions identified in each system with adequate precision.  This grading system is described in greater detail in Y09; an abbreviated explanation is given below.  

Lines contribute to the final confidence grade of a system if they meet the following four criteria:  

1.  The line has a significant detection.  For grading purposes, we require a lower limit of $\frac{W_{r}}{\sigma_{W_{r}}} \geq 4$, with rest-frame equivalent width, $W_{r}$, and respective error, $\sigma_{W_{r}}$.  The manual inspection of $\sim$1,000 spectra from earlier runs of the data has shown this limit to be the best compromise between increasing the ability to identify weak lines and minimizing the number of false detections.

2. The line is identified as a primary ion or neutral (Mg II, Fe II, Al II, Al III, C II, Si II, Mg I, C IV, Si IV, N V, or O VI).  These species have been chosen due to their high frequency of observation and reasonably high line strengths in the wavelength range covered by the SDSS.  

3. The line is not blended with any lines from another identified system in the same spectrum.  Absorption lines from systems at separate redshifts frequently overlap by chance, so our algorithm allows for the placement of a single observed line into more than one absorption system.  However, to conservatively rate the confidence of a line assigned to more than one distinct system (i.e., blended), we do not consider these identifications to be reliable.

4. The line is detected longward of the Ly-$\alpha$ forest.  Lines detected within the Ly-$\alpha$ forest are generally difficult to distinguish from abundant neutral hydrogen at other redshifts.  Therefore, lines detected shortward of Ly-$\alpha$ emission in each quasar spectrum are retained if matched to a system redshift, but these do not qualify as reliable and therefore do not contribute to the system grade.

5.  The line has a redshift offset from the system average $|z_{avg} - z_{line}| \leq 0.0013\times(1+z_{avg})$, corresponding to a velocity separation $<$400 km s$^{-1}$.  The average redshift of a system,$z_{avg}$, is determined through a two-step procedure.  An initial average system redshift is first calculated using all lines in a system that meet the first four criteria listed above.  Then, lines with redshift offsets from this average that exceed the allowed limit are removed and $z_{avg}$ is recalculated.

The number of lines which meet these five criteria are then summed for each system.  Systems with at least four lines with confident detections receive an ``A" grade; systems with 3 confident line identifications are classified as grade ``B".  Grade ``C" systems have two lines, which are generally Mg II doublets or C IV doublets.  This procedure is efficient, and the accuracy of grade A systems has been shown to be greater than 99\%.  Grade B, C, etc., systems are successively less reliable and require visual inspection for confirmation.  Additional grades for systems with lesser certainty exist (D, E, etc.), but these are ignored in our current analysis.  

We have examined how the ratio of the number of observed grade C systems relative to the number of systems qualifying as grade B, $N_{C}/N_{B}$, changes as a function of rest equivalent width of the 2796\AA $ $ line of Mg II.  We find  $N_{C}/N_{B}=15$ for $W_{r}^{\lambda2796}\leq0.3$\AA $ $, and this ratio reduces to a roughly constant $N_{C}/N_{B}\sim3$ for $W_{r}^{\lambda2796}\geq0.8$\AA $ $.  This behavior is an expected result of the correlation between detection confidence and absorber strength in a medium resolution survey.  Additionally, the weakest doublets are unlikely to have multiple weaker lines matched with 4$\sigma$ certainty due to the limiting resolution of the SDSS spectroscopy, and therefore weaker systems are inherently less likely to qualify as higher-grade detections.  Since grade B systems are considered in this sample to be confident identifications, we accept grade C systems to be equivalent to grade B where the ratio $N_{C}/N_{B}$ is approximately constant.  We determine that  Mg II doublets with as few as two lines meeting the above criteria exhibit the same approximate significance as  Mg II in grade B systems, provided that a minimum equivalent width of ($W_{r}^{\lambda2796}\geq0.8$\AA) is also required.  Therefore, grade A, B, \& C systems are included in the sample used for the cross-correlation analysis, as described in the following section.

A few noteworthy systematics have been found to persist in the production of the Y09 catalog.  The primary causes of error in system identification include poor continuum fitting around narrow emission features, doublet ratios deemed false due to the saturation of  Mg II and C IV lines, and strong absorption features identified within night sky emission bands or regions of the spectrum that are prone to atmospheric absorption.  An additional systematic is the tendency of the system-finding algorithm to catalog BALs as 5 or more systems blended together, which can produce a large quantity of closely-spaced narrow-line systems with low velocity separations from the background quasar.  By cross-checking with the SDSS DR5 BAL catalog (Gibson et al. 2009), these objects are flagged and later removed for the purpose of these clustering measurements.

\subsection{ Mg II Sample Refinement}

The entire Y09 SDSS DR5 QAL catalog contains 35,060  Mg II absorption systems with confidence grades of C or better over the observable redshift range $0.36<z<2.2$.  Within this sample, 7,702  Mg II systems are found to overlap in redshift-space with the LRGs used in this analysis, that is, $z_{abs}\leq0.8$.  In order to more simply quantify the effects of gravitational lensing present in our measurement of the Mg II -- LRG cross-correlation (discussed later in Section 5.2),  we require that the quasars in our sample do not overlap in redshift-space with the LRG population, having $z_{qso}>0.8$.  This additional requirement reduces our initial sample of  Mg II systems to 6,679.

Refining a sample of QALs with similar environmental origins is somewhat complicated due to the fact that quasar spectra frequently feature both absorption associated with the local quasar environment and intervening absorption, which is produced by gas and dust at lower redshifts physically unassociated with the background quasar.  Broad absorption lines (BALs; FWHM$>$2000 km s$^{-1}$), for example, have a well-established association with quasar outflows (e.g., Weymann et al. 1979; Yuan $\&$ Wills 2003; Richards 2006; Ganguly et al. 2007; Lundgren et al. 2007) and are thought to result from the intersection of the line of sight with wind radiated from the central accretion disk \citep{MC95,Elvis2000,Proga2000}.  BALs are observed in $\sim15\%$ of quasar spectra, a fraction that is presumably the result of the opening angle of this disk wind (Weymann et al. 1991; Hall et al. 2002; Reichard et al. 2003; Trump et al. 2006; Gibson et al. 2008).  As we are only interested in probing the environments of intervening  Mg II absorption, we exclude known BALs from this analysis, drawing from an SDSS DR5 BAL catalog mentioned in the previous section.

The vast majority of quasar absorption lines, however, are narrow (NALs; FWHM$\leq$500 km s$^{-1}$) and of more ambiguous origin.  In the case of NALs, differentiating between associated and intervening absorption poses an especially complex problem.  The observation of an excess of NALs in the proximity of the background quasar is well established (see, e.g., Weymann et al. 1979) and has been interpreted as a combination of absorption from galaxies in the neighborhood of the quasar and high-velocity gas contained in outflows originating from the quasar central engine (e.g., Weymann et al. 1979; Nestor et al. 2008).  The distribution of NALs with $v\lesssim3,000$ km s$^{-1}$ measured in the quasar rest frame can generally be explained by a combination of gravitationally bound gas within the local quasar environment and supernovae-driven winds, whereas outflows from the central engine of the quasar provide an explanation for the excess of NALs at significantly higher velocities.

Wild et al. (2008) report that the velocity distribution of CIV NALs associated with the quasar extends to $\beta\sim0.04$\footnote[1]{Measurements of absorber velocities in the quasar rest frame are commonly described using the dimensionless quantity, $\beta$, where
\begin{equation}
\beta=\frac{v}{c}= \frac{\left[(1+z_{qso})/(1+z_{abs})\right]^{2}-1}{\left[(1+z_{qso})/(1+z_{abs})\right]^{2}+1}
\end{equation}
} and observe an excess of associated Mg II absorption primarily contained within $\beta<0.02$.  These recent results are supported by the finding that $\sim40\%$ of all W$_{r}^{\lambda1548}\geq 0.3$\AA $ $ C IV QALs within $\beta<0.04$ can be attributed to quasar outflows (Nestor et al. 2008).   Mg II NALs with velocities of $\beta<0.01$ have also been found to exhibit higher apparent ionization, extinction, and mean rest equivalent width, compared to those with  $\beta>0.01$ (Vanden Berk et al. 2008).  Together, these findings provide strong physical evidence that a significant fraction of low-$\beta$ NALs may be intrinsic to quasar outflows.  

Since BALs are known to be associated with quasar outflows and have been observed at velocities as great as $\beta=0.22$ \citep{Foltz83}, it remains uncertain whether a velocity cut of $\beta\sim0.04$ is sufficient for cleanly distinguishing between intervening and associated NALs.   Richards et al. (1999) reported an excess of NALs extending to $\beta\sim0.22$ and peaking at $\beta\sim0.1$ and observed differences in the velocity distribution of C IV NALs that were correlated with the steepness of the spectrum and radio-loudness of the host quasar.  Wild et al. (2008) similarly observed a significantly greater number of associated Mg II absorption systems in radio-loud quasars, compared to their radio-quiet counterparts.  By requiring that all  Mg II systems have $\beta>0.2$, we ensure that our sample of  Mg II systems is unbiased by the radio-type and orientation of the background quasar and free of contamination from associated absorption. This cut reduces the  Mg II sample size to 5,931.

\begin{figure}[t!]
\begin{center}
\plotone{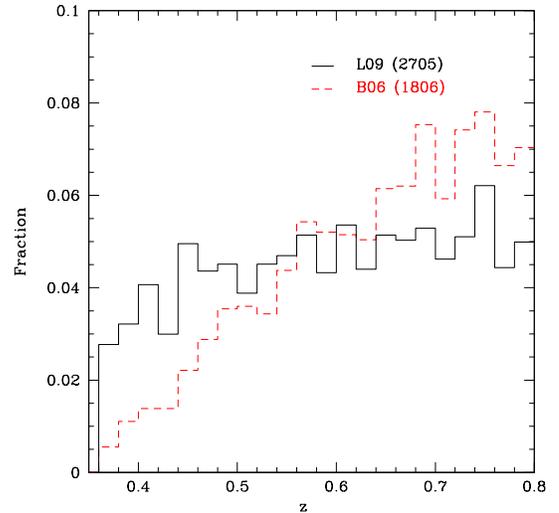}
\caption{The fractional redshift distribution of DR5  Mg II systems used in this work, compared to the sample of B06 (shown in red).  The difference in the steepness of the two distributions results from the removal of z$_{qso}\leq$0.8 quasars and $\beta\leq$0.2  Mg II systems in the Y09 sample.}
\end{center}
\end{figure}

We have introduced a few additional requirements to increase the confidence of detection for the remaining systems. The selection function for  Mg II in the SDSS sample depends on both equivalent width and the magnitude of the background quasar.  Brighter background sources increase the S/N of observations in the SDSS, and as a result we find that narrow features (W$_{r}\leq1.0$\AA) are more easily detected in the spectra of brighter quasars and infact comprise the majority of all detected systems.  In fainter quasars, we may still detect stronger lines, but the fraction of detected narrow features is greatly reduced. To minimize this selection effect we do not include lines with $W_{r}^{\lambda2796}<0.8$\AA $ $ in this analysis, reducing the sample to 4,330.  Since the number distribution of absorption systems rises with decreasing equivalent width, and the weakest lines become increasingly difficult to detect at faint magnitudes, we also exclude the faintest quasars (those with $i$-band apparent magnitudes $m_{i}>$19.5) from our sample in order to more adequately reflect the true equivalent width distribution of absorbers.  This magnitude cut removes an additional 1,106 systems.  Masking the sample to remove objects in areas of the survey with seeing greater than 1.5$\arcsec$ and high Galactic reddening, $A_{r}>0.2$, (using the same method as Ross et al. 2008) removes another 519 systems.

\begin{figure}[t!]
\begin{center}
\plotone{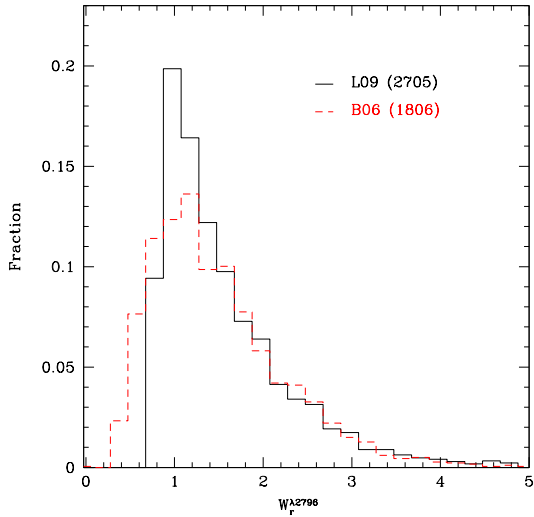}
\caption{The fractional rest-frame equivalent width distribution of  Mg II lines used in this work (black; hereafter, L09), compared to the distribution of lines in B06 (red). The peak of the distribution appears higher for the L09 sample, due to our imposed lower-limit of 0.8\AA.}
\end{center}
\end{figure}

The final sample of  Mg II used in our calculation of the Mg II -- LRG cross-correlation includes 2,705 absorption systems  covering a redshift range from $0.36\leq$ z$_{abs}\leq0.8$, with a mean redshift of z$_{abs}\sim0.6$.  The fractional redshift distribution is presented in Figure 1 with the distribution of the sample reported by B06 overplotted for comparison.  We find a much flatter distribution for the sample used in this analysis.  This comparative paucity of high-redshift absorbers in the sample we have compiled is due to our removal of Mg II systems with $\beta\leq0.2$, which preferentially decreases the  number of  Mg II systems on the high-redshift end of the sample and more accurately represents the distribution of intervening absorbers (see Section 5.1 for a detailed discussion of this effect).  We present the fractional distribution of our sample as a function of the rest equivalent width of the 2796\AA $ $ line in Figure 2, where again we show the distribution of B06 for comparison. The rest equivalent widths of the 2796\AA $ $  Mg II absorption in our sample range from 0.8\AA$\leq W_{r}^{\lambda2796}<$5.0\AA, with a mean of W$_{r}^{\lambda2796}\simeq1.4$\AA $ $.  We find that the equivalent width distribution of our sample agrees quite well with that of B06, with the exception of the peak values, which differ as a result of the lower limit we have chosen for $W_{r}^{\lambda2796}$.   

For completeness, we have included an electronic catalog of Mg II absorbers used for analysis in this work.  This sample of 3,469 systems has been constructed according to the requirements presented above, with the exception of the limits on the quasar redshift and absorber velocity.  We have chosen to include the low-$\beta$ systems in the electronic table, since we explore the effects of these systems on our measurement of the Mg II -- LRG cross-correlation in Section 5.1.  An excerpt from this catalog is presented in Table 1.  The complete criteria used to extract these data from the Y09 catalog are the following:  grade$\geq$C; $W_{r}^{\lambda2796}\geq0.8$\AA; $m_{i}\leq19.5$; $z_{abs}\leq0.8$; seeing$<$1.5\arcsec; reddening$<$0.2.

\begin{figure}[t!]
\begin{center}
\plotone{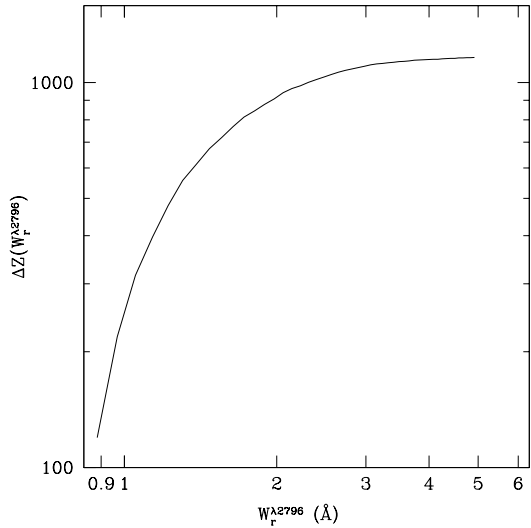}
\caption{The redshift path covered by the sample of 2,705 Mg II absorbers used in this work,  $\Delta$Z(W$_{r}^{\lambda2796}$), shown as a function of W$_{r}^{\lambda2796}$.}
\end{center}
\end{figure}

The method we employ in this work to measure the clustering amplitude of these absorbers is ultimately biased by the redshift distribution of the samples, so it is critical that both the galaxy and absorber reshift distributions are well understood.  A number of measurements have been made of the redshift number density of  Mg II within the redshift range of this study (Tytler et al. 1987; Sargent et al. 1988; Caulet 1989; Steidel \& Sargent 1992; Churchill et al. 1999; Ellison et al. 2004).  Most recently, Nestor et al. (2005) measured the evolution of the  Mg II number density using  $\sim$1300 systems extracted from the SDSS Early Data Release.  We have measured the redshift number density, $\partial$N/$\partial$z, of  Mg II in this work as a function of redshift for a series of slices in equivalent width.  The incidence of lines in an interval of  $W_{r}^{\lambda2796}$ over a particular redshift path is defined as:
\begin{equation}
\frac{\partial N}{\partial z} = \displaystyle\sum_{i}\frac{1}{\Delta Z(W_{r}^{\lambda2796})},     
\end{equation}
where
\begin{equation}
\Delta Z(W_{r}^{\lambda2796})= \displaystyle\int^{z_{max}}_{z_{min}} \displaystyle\sum_{i=1}^{N_{spec}} g_{i} (W_{r}^{\lambda2796}, z) dz
\end{equation}
denotes the total redshift path of the survey for some equivalent width interval.  Here, $g_{i} (W_{r}^{\lambda2796}, z)=1$ for every detection of a line with $W_{r}^{lim}\leq W_{r}^{\lambda2796}$ and equals zero otherwise (Lanzetta et al. 1987). In determining the observable redshift range, $dz$, for each quasar sightline we have required that Mg II be observable in SDSS wavelengths outside the Ly$\alpha$ forest and have a velocity greater than 60,000 km s$^{-1}$ in the quasar rest frame.  The variance of the redshift number density is given as:
\begin{equation}
\sigma^{2}_{\partial N / \partial z} = \displaystyle\sum_{i}\left[\frac{1}{\Delta Z(W_{r}^{\lambda2796})}\right]^{2}.
\end{equation}
The  redshift path (Eq. 8) summed over the total redshift range of this sample ($0.36\leq z \leq0.8$) is shown as a function of $W_{r}^{\lambda2796}$ in Figure 3.

\begin{figure}[t!]
\begin{center}
\plotone{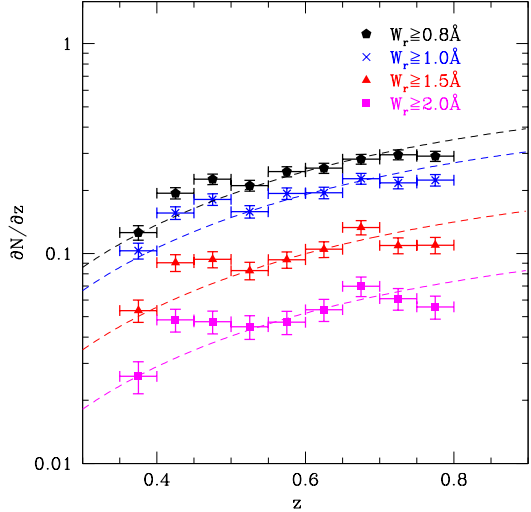}
\caption{The evolution of the redshift number density, $\partial$N/$\partial$z, for equivalent-width-limited samples of  Mg II used in this work with 1$\sigma$ errors.  Dashed curves indicate the $\chi^{2}$ best fits to a non-evolving redshift number density.  These data agree with the SDSS EDR measurements of Nestor et al. (2005).  The samples are marginally consistent with no evolution over the redshift and equivalent width ranges examined in this work, although some deviations from the non-evolution curves are noticeable within the redshift range $0.4<z<0.5$ and as the distribution approaches z$\sim$0.8.}
\end{center}
\end{figure}

\begin{figure}[h!]
\begin{center}
\plotone{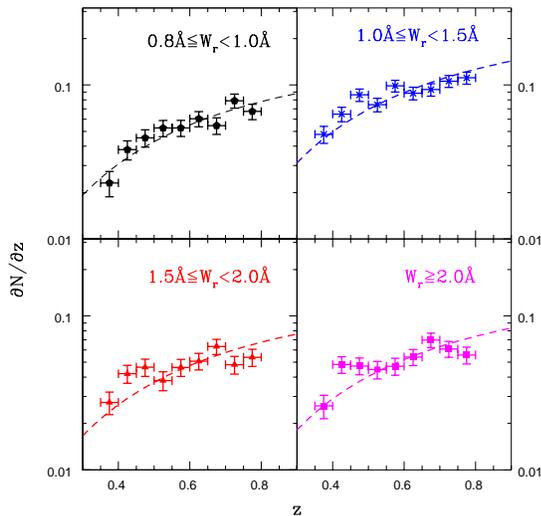}
\caption{The same measurements as in the previous figure, with narrower slices in rest equivalent width. The lowest equivalent width sample (0.8\AA$<W_{r}^{\lambda2796}<1.0$\AA) is consistent with a non-evolving redshift number density over the range $0.4\lesssim z \lesssim 0.8$ to 1$\sigma$.  However, we find increasingly more significant deviations from the NEC curves for samples with larger mean equivalent widths.  The flattening of the measured $\partial$N/$\partial$z relative to the best-fit NEC curves for the larger equivalent width samples (particularly $W_{r}^{\lambda2796}>2.0$\AA) suggests that these populations exhibit significant redshift number density evolution over $0.4\lesssim z \lesssim 0.8$.}
\end{center}
\end{figure}

We have measured the redshift number density as a function of redshift for a number of equivalent width-limited samples.  These measurements are provided in Table 2 and presented in Figure 4 with no-evolution curves (NECs) overplotted.  The NECs trace a constant comoving absorber number density as a function of redshift and have been normalized for each sample to minimize the $\chi^{2}$ best fit to the measured $\partial N / \partial z$ over the redshift range $0.36\leq z\leq0.8$.  The amplitudes of the best-fit NECs agree within 1$\sigma$ with those measured by Nestor et al. (2005).   Additionally, the greater number of absorbers included in this work allows for higher resolution measurements of the evolution of the $\partial N / \partial z$ than previous analyses.  

The data in Figure 4 show 1$\sigma$ divergence from the best-fit to the NECs over the redshift range $0.4\lesssim z\lesssim0.5$, providing evidence for redshift evolution of these samples.  In Figure 5 we investigate the evolution of the redshift number density with narrower cuts on equivalent width.  The measurements are additionally presented in Table 3.  We find that the 0.8\AA$\leq W_{r}^{\lambda2796}<1.0$\AA $ $ sample fits well to the 1$\sigma$ NEC best-fit curve.  However, we find that the minimum $\chi^{2}$ value of these fits to a non-evolving population increases with increasing equivalent width.  This implies that these larger equivalent width samples (particularly W$_{r}^{\lambda2796}\geq2.0$\AA) are evolving significantly over the redshift range examined in this work.  The observed differences in evolutionary behavior of the various equivalent width samples has further implications with regard to their associated host galaxies, which we address in Section 5.2.

\subsection{LRG Sample Selection and Random Catalogs}

Due to the high precision of spectroscopic redshifts, one would prefer to calculate the absorber-LRG cross-correlation using two spectroscopically observed samples.  However, the design of the SDSS spectrograph, paired with the method of targeting priorities and the infrequency of repeat observations, complicates the SDSS spectroscopic selection function and limits our ability to measure small-scale cross-correlations between spectroscopic samples in the SDSS (e.g., mechanical constraints do not permit objects within 55$\arcsec$ to be observed on the same spectroscopic plate).  Photometric catalogs provide an elegant solution to this problem.  The work of Eisenstein (2001) and others (e.g., Brown et al. 2003; Zehavi et al. 2005a; Eisenstein 2005b) have demonstrated that the strong clustering nature of LRGs, in particular, can provide significant signal to overcome the uncertainties in photometric redshift estimates and make extremely sensitive angular correlation measurements.

The sample of LRGs in this work have been drawn from the SDSS DR5 photometric catalog created by Ross et al. (2008), which followed the color selection and redshift estimation prescriptions of Collister et al. (2007).   The selection algorithm employed by Collister et al (2007) to produce the MegaZ-LRG photometric catalog of more than one million LRGs over a redshift range $0.4<z<0.7$ selected from the Fourth Data Release of the SDSS is provided below:
\begin{equation}
i_{fiber}<21.4
\end{equation}
\begin{equation}
17.5<i_{deV}<20.0
\end{equation}
\begin{equation}
0.5< g - r < 3
\end{equation}
\begin{equation}
r - i < 2
\end{equation}
\begin{equation}
c_{\parallel}\equiv0.7(g-r) + 1.2(r-i-0.18) > 1.6
\end{equation}
\begin{equation}
d_{\perp}\equiv(r-i)-(g-r)/8.0>0.5
\end{equation}
Effective star-galaxy separation is achieved by requiring:
\begin{equation}
i_{psf}-i_{model} > 0.2(21.0-i_{deV})
\end{equation}
\begin{equation}
r_{i_{deV}}> 0.2\arcsec
\end{equation}
where $i_{fiber}$ represents the $i$-band flux contained within the aperture of a SDSS spectroscopic fiber; $i_{dev}$ gives the $i$-band magnitude resulting from a best fit to a de Vaucouleurs profile with radius $r_{i_{deV}}$; $i_{psf}$ represents the magnitude determined from a best fit of the PSF at the galaxy position; and $i_{model}$ is the $i$-band magnitude resulting from a best fit to a de Vaucouleurs or exponential profile in the $r$-band.  Objects were further required to have SDSS flags indicating nchild=0, not SATURATED in any band, and detected and not NO PETRO in bands $r$ or $i$.

Redshifts and corresponding errors were assigned to photometric targets meeting these requirements by Ross et al. (2008) using the ANNz software (Firth et al. 2003) and the Two-Degree Field-SDSS LRG and QSO (2SLAQ) spectroscopic LRG catalog (Cannon et al. 2006).  The resulting catalog consists of nearly 1.7 million LRGs with a median photometric redshift of z=0.52.  As in Collister et al. (2007) the photometric redshifts in the sample have an rms photometric accuracy of $\sigma_{z}=$0.049, and the estimated stellar contamination is less than 2\% (as shown by Ross et al. 2008).  The full catalog of LRGs has been masked using the same seeing and Galactic reddening criteria applied to the QAL sample.  With these masks applied, 1,495,604 LRGs remain for use in this analysis, providing a much larger and higher-redshift sample than the available spectroscopic data in the SDSS DR5.  The redshift distribution of the entire LRG sample is shown in Figure 6.  

This DR5 LRG sample contains 12,292 LRGs with z$>$0.7, which have been shown by Collister et al. (2007) to have photometric redshift estimates with variance $\sigma_z$=0.054.  Although LRGs with z$>$0.7 constitute less than 1\% of the overall sample, we have included these in the Mg II -- LRG cross-correlation since the LRG photometric redshift errors are still quite good, and this region overlaps with $\sim$25\% of the absorbers available in our analysis.  However, the number of MgII -- LRG pairs at z$>$0.7 contributes only 1.3\% to the overall number of Mg II -- LRG pairs with separations $<$40 h$^{-1}$Mpc in our sample, so the effects of the highest redshift objects in the cross-correlation of our full sample should be nearly negligible.

We have constructed a random LRG photometric redshift catalog by choosing random coordinates in the area of sky covered by the SDSS DR5 that also satisfy the seeing and Galactic reddening requirements of Ross et al. (2008).   We produced over 180 million random coordinates for the analysis, outnumbering the real photometric data by a factor of 124.  Redshifts from the photometric LRG catalog were then shuffled and randomly assigned to the catalog of masked random coordinates.  By sampling the redshift distribution of the photometric data in this way, we preserve the same redshift distribution for all real and random LRG data.

\section{Correlation Analysis}

The spatial correlation function, $\xi(r)$, is typically used to quantify clustering amplitudes, since it describes the probability above random of detecting a target object within a physical distance $r$ of a particular source (Totsuji \& Kihara 1969; Peebles 1974).  The distribution of baryonic matter is expected to trace the clustering of the underlying dark matter, which scales as a function of dark matter halo mass, $M$ (see, e.g., Mo, Peacock \& Xia 1993; Mo \& White 2002).  Thus, in a linear bias model, the correlation function of baryonic matter is related to the dark matter correlation function, $\xi_{DM}$, according to a mean bias, $\bar{b}(M)$.  We explicitly consider the galaxy auto-correlation function, $\xi_{gg}(r)$, and absorber-galaxy cross-correlation function, $\xi_{ag}(r)$.  The relationship of these functions to the underlying dark matter distribution are given below:
\begin{equation}
\xi_{gg}(r) = \bar{b}^{2}(M_{g})\xi_{DM}(r)
\end{equation}
\begin{equation}
\xi_{ag}(r) = \bar{b}(M_{a})\bar{b}(M_{g})\xi_{DM}(r)
\end{equation}

\begin{figure}[t!]
\begin{center}
\plotone{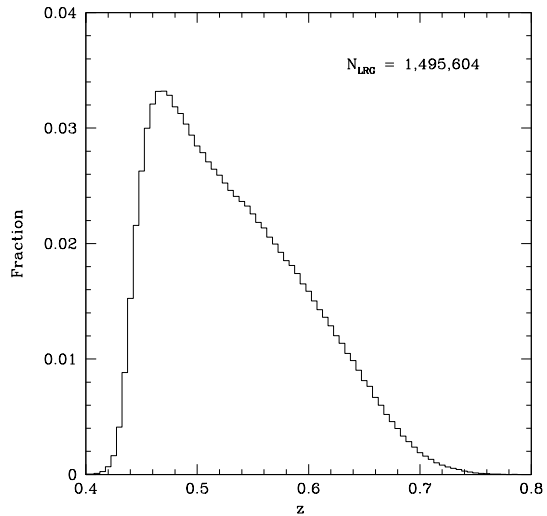}
\caption{The fractional photometric redshift distribution of LRGs used in this work.}
\end{center}
\end{figure}

As explained in detail by Bouch{\'e} et al. (2004, 2005), for galaxies with a well-known halo mass ($M_{g}$) one can quite simply estimate a typical absorber dark matter halo mass by exploiting the symmetry in the above relationships.  Provided the same galaxy sample is used for each measurement,  the ratio of the absorber-galaxy cross-correlation and the galaxy auto-correlation is equivalent to the ratio of the absorber and galaxy dark matter biases.   For a galaxy sample with a precisely measured dark matter bias, such as the LRGs, we need only measure the auto- and cross-correlation functions in order to estimate the bias of the absorber sample, as shown below:
\begin{equation}
\bar{b}(M_{a})=\frac{\xi_{ag}}{\xi_{gg}}\bar{b}(M_{g})
\end{equation}

\subsection { Mg II -- LRG Cross-Correlation}

A spatial cross-correlation produces precise measurements of cross-correlations among samples with well-determined redshifts.  However, our photometrically-determined LRG redshifts include errors as large as $\sigma_{z}=0.05$, which can be a significant source of scatter for spatial cross-correlations.  To obtain the most accurate measurement of the  Mg II -- LRG clustering, we instead calculate the projected cross-correlation, $\omega(r_{\theta})$, which, as shown in B06, can be used interchangeably with the spatial cross-correlation to return a precise measurement of the dark matter bias of the absorbers relative to the LRGs.  To calculate $\omega(r_{\theta})$ we de-project the LRGs contained within redshift slices of width $|z_{abs} - z_{gal}| \leq 0.05$ around each individual absorber and measure the angular cross-correlation in each case.  As demonstrated in B06,  this choice for the widths of the slices adequately accounts for the uncertainty in the LRG photometric redshifts and does not bias the ultimate measurement of the  Mg II dark matter halo mass, provided that the auto-correlation of the LRGs symmetrically applies the same redshift bin widths.  

In concordance with B06, we calculate the projected cross-correlation using the estimator employed by Adelberger et al. (2003): 
\begin{equation}
\omega_{ag}(r_{\theta})=\frac{AG}{AR} - 1.0
\end{equation}
where $r_{\theta}$ = $\chi \theta$, with $\chi$ representing the angular diameter distance to the absorber redshift, measured in comoving h$^{-1}$ Mpc, and $\theta$ denotes the subtended angle.  $AG$ represents the number of absorber-galaxy pairs within the annulus defined by $r_{\theta}-dr_{\theta}/2 < r_{\theta} < r_{\theta}+dr_{\theta}/2$, and $AR$  represents the number of absorber-random galaxy pairs within that same annulus normalized by the ratio of the total real galaxies $N_{G}$ to the total random galaxies $N_{R}$ determined independently for each of the jackknife regions described below.  

We measure the full covariance error matrices according to the jackknife resampling method (e.g., Efron \& Gong 1983) by partitioning our data into 10 regions defined by evenly-populated strips of sky (Zehavi et al. 2002; Myers et al. 2005).   We calculate $\omega(r_{\theta})$ 10 times, each time removing a different region from the analysis.   The statistical covariance of $\omega(r_{\theta})$ for each partition, COV$_{ij}$, is then calculated according to the following estimator:
\begin{equation}
COV_{ij}=\left(\frac{N-1}{N}\right)\sum_{l=1}^{N} [\omega_{l}(r_{\theta_{i}}) - \bar{\omega}(r_{\theta_{i}})][(\omega_{l}(r_{\theta_{j}}) - \bar{\omega}(r_{\theta_{j}})] 
\end{equation}
where $\bar{\omega}(r_{\theta_{i}})$ is the mean value of $\omega_{l}(r_{\theta_{i}})$ measured for all N=10 regions.  

\subsection {LRG Auto-Correlation}

In order to calculate the ratio of the Mg II -- LRG cross-correlation and the LRG auto-correlation amplitudes we must account for the fact that we are cross-correlating photometric and spectroscopic samples.  Therefore, for this analysis, we must use an estimator that is symmetric to Equation (21) when calculating the auto-correlation of the photometrically selected LRGs (for a detailed explanation see Bouch{\'e} et al. 2005, appendix A):
\begin{equation}
\omega_{gg}(r_{\theta})=\frac{GG}{GR} - 1.0
\end{equation}
where $GG$ represents the number of galaxy-galaxy pairs within the annulus defined by $r_{\theta}-dr_{\theta}/2 < r_{\theta} < r_{\theta}+dr_{\theta}/2$, and $GR$ represents the number of galaxy-random galaxy pairs within that same annulus, normalized by the ratio of the total number of real galaxies $N_{G}$ to the total number of random galaxies $N_{R}$ in the same jackknife partition as each LRG.  In the same manner as the  Mg II -- LRG calculation, we de-project the angular auto-correlation around each absorber, including only LRGs that lie in the range $|z_{abs} - z_{LRG}| \leq 0.05$.

\section{Results}

\begin{figure}[t!]
\begin{center}
\plotone{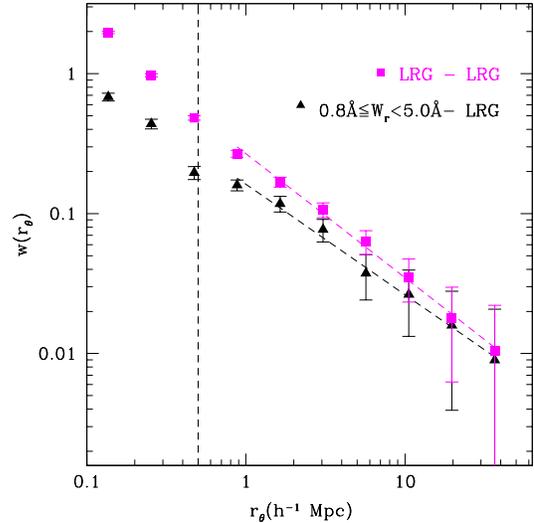}
\caption{The projected cross-correlation of  Mg II with LRGs, compared to the projected auto-correlation of LRGs.  The  Mg II -- LRG measurements have been shifted slightly along the x-axis for easier viewing.  This sample of absorption lines have been selected with the following criteria: $\beta>$0.2, 0.36$\leq$ z$_{abs} \leq$0.8, 0.8$\leq$z$_{qso}$$\leq$4.5, $m_{i}\leq$19.5, and 0.8\AA$\leq$ W$_{r}^{\lambda2796}$$\leq5.0$\AA.  Error bars include consideration for the effects of cosmic variance and have been calculated with standard Jackknife resampling of 10 regions.  The best power-law fit for each function is overplotted over the range of data included for fitting, and the respective fit parameters are provided in Table 6.  The vertical dashed line denotes the scale below which weak lensing is expected to affect this measurement (see Section 5.2).}
\end{center}
\end{figure}

In Figure 7 we present measurements of the projected  Mg II -- LRG cross-correlation and the LRG auto-correlation.  These measurements are also provided in Table 4. The dashed vertical line at 0.5 h$^{-1}$ Mpc (also shown in subsequent correlation plots) denotes the approximate scale at which weak lensing  (see Section 5.2 for details) is likely to have affected our measurements.  

To first order, spatial cross-correlations are best described by fits to a power-law (Totsuji \& Kihara 1969; Peebles 1974).  The LRG two-point correlation is well-modeled by this form, although deviations have been found in high precision measurements (e.g., Berlind et al. 2003; Magliocchetti \& Porciani 2003; Maller et al. 2005; Zehavi et al. 2004;  Zheng 2004).  A known break exists in the power-law slope of the LRG auto-correlation at $\sim$1 h$^{-1}$ Mpc (see, e.g., Zehavi et al. 2005), which can be explained by modeling the changing contributions of one- and two-halo terms in a multiple dark matter halo regime.  We confirm this feature in our measurement of the LRG auto-correlation and only include measurements on larger scales when fitting our results.  To calculate the best-fit power law (of the form $\widehat{\omega}=A r_{\theta}^{\gamma}$) to each correlation function, we minimize:
\begin{equation}
\chi^{2} \equiv \frac{1}{N_{dof}}[\omega - \widehat{\omega}]^{T}COV^{-1}[\omega - \widehat{\omega}]
\end{equation}
over the range 1$-$40 h$^{-1}$ Mpc.  We then measure the amplitude, $A$, of each fit at $r_{\theta}=1$ h$^{-1}$ Mpc for the  Mg II -- LRG cross-correlation and the LRG auto-correlation.  The ratio of these amplitudes, $a = A_{ag}/A_{gg}$, provides an approximate measurement of the bias ratio, $b_{ag}/b_{gg}$, of the two species (shown in Equation 20):
\begin{equation}
a \equiv \frac{\widehat{\omega}_{ag}}{\widehat{\omega}_{gg}}
\end{equation}

We find the best-fit parameters for a power law fit to the LRG auto-correlation to be $A_{gg}=0.268\pm0.052$ and $\gamma_{gg} = -0.886\pm0.033$.  The  Mg II -- LRG cross-correlation returns best-fit parameters $A_{ag}=0.162\pm0.013$ and $\gamma_{ag}=-0.789\pm0.035$; therefore the amplitude ratio for our entire  Mg II sample is $a=0.605\pm0.13$.  The fit parameters used to obtain this ratio  are provided in Table 6.  As noted in B06, Gaussian errors on the photometric redshifts in an LRG sample should propagate into the measurement of the amplitude ratio above, causing this ratio to over-estimate the true dark matter bias ratio, $b_{ag}/b_{gg}$.  Calculations by B06 approximate the surplus in this measurement to be 25$\pm$10 percent for their sample of LRGs with average redshift errors of $\sigma_{z_{phot}}=0.1$.   As the LRG redshift distribution in our DR5 sample is better established and has a factor of 2 improvement in photometric redshift errors compared to the LRG sample of B06, we do not apply this correction factor and take $\omega_{ag}/\omega_{gg}$ to be an accurate estimate of the bias ratio $b_{ag}/b_{gg}$.  We provide a thorough discussion of the effect of decreased photometric errors in Section 5.1.  The value we recover for this ratio, 0.605$\pm$0.13, is consistent with the bias ratio expected for late- to early-type galaxies, $b_{late}/b_{early}\sim0.7$ (Bouch{\'e} et al. 2004).

Ross et al. (2008) measured the dark matter bias of the same LRG sample we use to be $b=1.82\pm0.02$ at the mean redshift of intervening absorbers in this work (z=0.61) and with $\sigma_{8} = 0.79$.  The bias of halos as a function of halo mass can be determined following an ellipsoidal collapse model (e.g., Sheth et al. 2001).  Assuming the same cosmology as Ross et al. (2008) -- a flat Universe with $\sigma_{8} = 0.79$, $\Omega_{m} = 0.238$, $h =0.73$, and $\Gamma = 0.135$ (where $\Gamma$ is the shape parameter defined by Eisenstein \& Hu 1998) -- a dark matter halo bias of 1.82 corresponds to a dark matter halo mass of log$M_{h}$(M$_{\sun}h^{-1}$)=13.4 for halos at z=0.61.  Taking this to be the mass of the dark matter haloes hosting LRGs in our sample, we can approximate the dark matter halo masses associated with the average Mg II absorber using Equation 20.  The amplitude we measure of the Mg II absorbers relative to the LRGs, $b_{ag}/b_{gg}=0.605\pm0.13$, returns an absorber dark matter halo bias 1.101$\pm$0.24, corresponding to a halo mass of log $M_{h}$($M_{\sun}h^{-1}$)=12.11$\pm^{0.63}_{1.68}$, which is $\sim$20 times smaller than the LRG dark matter halo mass and consistent with the value obtained by B06 (log $M_{h}$($M_{\sun}h^{-1}$)=11.94$\pm^{0.39}_{0.40}$).

\begin{figure}[t!]
\begin{center}
\plotone{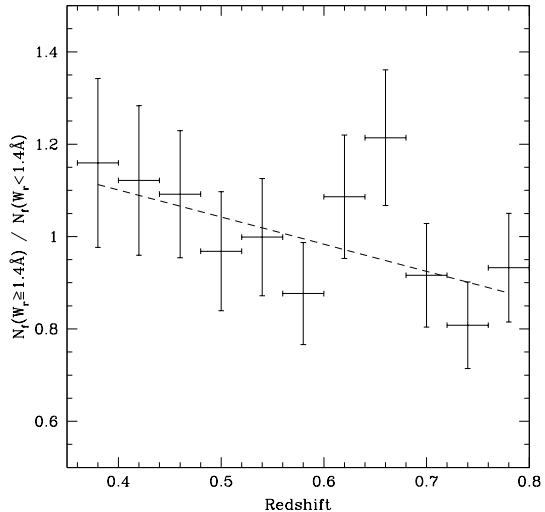}
\caption{ The ratio of the fractional number of W$_{r}^{\lambda2796}\geq1.4$\AA $ $ systems versus the fractional number of 0.8\AA $ $ $\leq$W$_{r}^{\lambda2796}<1.4$\AA $ $ systems in this analysis, shown as a function of redshift with Poisson errors.  A linear weighted least squares best fit function of  the form $y = ax + b$ with $a=-0.586\pm0.32$ and $b=1.335\pm0.197$ is overplotted.}
\end{center}
\end{figure}

\begin{figure}[t!]
\begin{center}
\plotone{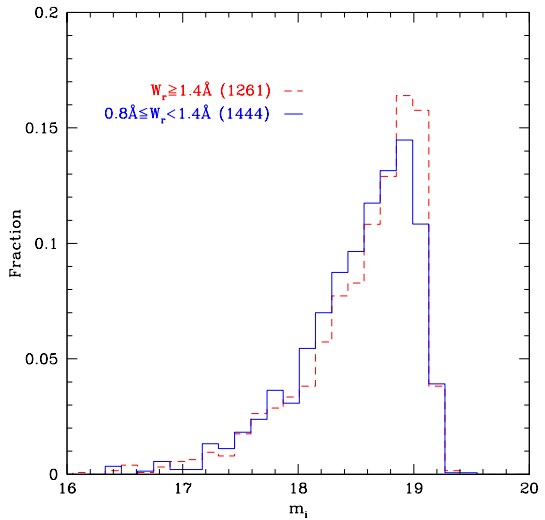}
\caption{The fractional $i$-band apparent magnitude distribution for the two primary equivalent width samples of  Mg II used in this analysis.  Due to the little difference in these distributions, we may expect any weak lensing to affect our two primary equivalent width samples similarly.}
\end{center}
\end{figure}

\begin{figure}[t!]
\begin{center}
\plotone{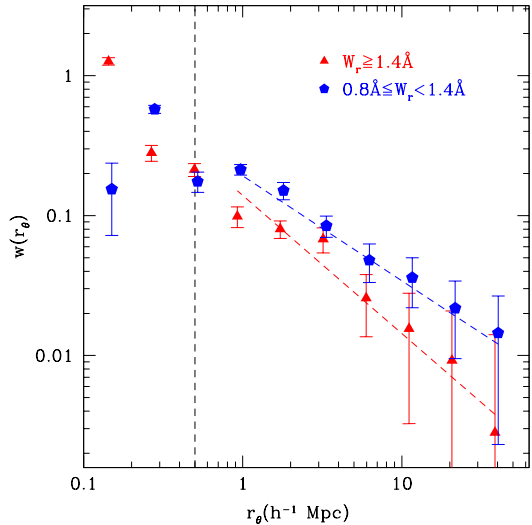}
\caption{The projected cross-correlation of two samples of DR5  Mg II absorbers with DR5 LRGs. The large ($W_{r}^{\lambda2796}\geq1.4$\AA) and small ($0.8$\AA$\geq W_{r}^{\lambda2796}<1.4$\AA) equivalent width samples include 1,261 and 1,444 lines, respectively.  The $0.8$\AA$\geq W_{r}^{\lambda2796}<1.4$\AA$ $ cross-correlation measurements have been shifted slightly along the x-axis for easier viewing.  The best power-law fit for each sample is overplotted over the range of data included for fitting, and the respective fit parameters are provided in Table 6.  The vertical dashed line denotes the scale below which weak lensing is expected to affect this measurement (see Section 5.2).}
\end{center}
\end{figure}

We additionally cross-correlate two samples of  Mg II divided according to the rest equivalent width of the 2796\AA $ $ line, to investigate the relationship of equivalent width and clustering amplitude.  Our sample of  Mg II contains 1,261 absorbers with $W_{r}\geq1.4$\AA $ $ and 1,444 with $0.8$\AA$\leq W_{r}<$1.4\AA $ $.  The respective mean redshift for each sample is: z=0.591 and z=0.600.  In Figure 8 we present the ratio of the fractional redshift distributions of these two sub-samples with Poisson errors and a minimum $\chi^{2}$ linear fit overplotted.  A best fit to a linear function of the form $y = mx + k$ returns $m=-0.586\pm0.32$ and $k=1.335\pm0.20$, indicating that the ratio of the fractional redshift distribution of the $W_{r}\geq1.4$\AA $ $ sample to that of the $0.8$\AA$\leq W_{r}<$1.4\AA $ $ sample decreases slightly with increasing redshift.  This is not surprising, given the width-dependent evolution of  redshift number density shown in Figure 5.  Significant differences in the redshift distribution of  Mg II sub-samples complicate the comparison of the correlation amplitudes;  however, though we see a redshift dependency of the ratio over the entire sample, the trend is not significant when the absorber sample is restricted to z$<$0.7, the region containing 99\% of our LRG sample.  A minimum $\chi^{2}$ linear fit to the ratio of $W_{r}\geq1.4$\AA $ $ and $0.8$\AA$\leq W_{r}<$1.4\AA $ $ absorbers with z$<$0.7 returns a slope of $m=-0.121\pm0.32$, which is consistent with finding no difference in the redshift distribution of the two equivalent width samples within the redshift range contributing the bulk of the cross-correlation signal.

In Figure 9 we present the fractional $i$-band apparent magnitude distribution for quasars hosting  Mg II absorbers in each of the two equivalent width sub--samples and show that these are similar as well.  The importance of matching magnitude distributions among the two absorber samples is discussed in more detail when we consider the potential biasing effects of weak lensing in Section 5.3.

In Figure 10 we present the cross-correlation of these two equivalent width-cut samples with the DR5 LRGs.  These measurements are also provided in Table 5.  We find, in agreement with B06, that the 0.8\AA$\leq W_{r}<$1.4\AA $ $ sample exhibits marginally stronger clustering compared to the sample of W$_{r}$$\ge$1.4\AA $ $  Mg II systems.  Fitting each function to a power law produces relative amplitudes of $a_{1}=0.52\pm0.12$ for the larger equivalent width sample and $a_{2}=0.72\pm0.18$ for the smaller equivalent width sample.  From these measurements we extract a dark matter halo bias of $b_{1}=0.94\pm0.22$ for the larger equivalent width sample and $b_{2}=1.31\pm0.32$ for the smaller equivalent width absorbers. Using the same formalism previously described, we find approximate dark matter halo masses to be log$M_{h}(M_{\sun}h^{-1})=11.29\pm^{0.36}_{0.62}$ for absorbers with W$_{r}$$\ge$1.4\AA $ $  and log$M_{h}(M_{\sun}h^{-1})=12.70\pm^{0.53}_{1.16}$ for the sample with 0.8\AA$\leq W_{r}<$1.4\AA $ $.  These measurements indicate that haloes hosting Mg II absorption with 0.8\AA$\leq W_{r}<$1.4\AA $ $ are typically $\sim$25 times more massive than haloes hosing absorbers with W$_{r}$$\ge$1.4\AA $ $.  Furthermore, the dark matter halo mass of the 0.8\AA$\leq W_{r}<$1.4\AA $ $ are consistent, within the measured error, with the average mass of LRGs in this redshift range.

In order to explore the apparent anti-correlation of equivalent width and clustering amplitude in greater detail, we further subdivided the  Mg II sample into four smaller samples:  0.8\AA$\leq W_{r}<$1.0\AA $ $, 1.0\AA$\leq W_{r}<$1.5\AA $ $, 1.5\AA$\leq W_{r}<$2.0\AA $ $, and 2.0\AA$\leq W_{r}<$5.0\AA $ $, which respectively contain 590, 1,002, 541, and 595  Mg II systems.  We again measured the cross-correlation of each sample with the LRGs and calculated the best-fit power law on scales 1$-$40 h$^{-1}$Mpc for each.  The best-fit parameters for each sample are contained in Table 6, and a plot of the  Mg II -- LRG clustering amplitude ratio as a function of rest equivalent width is presented in Figure 11 along with the published values from B06.

\begin{figure}[t!]
\begin{center}
\plotone{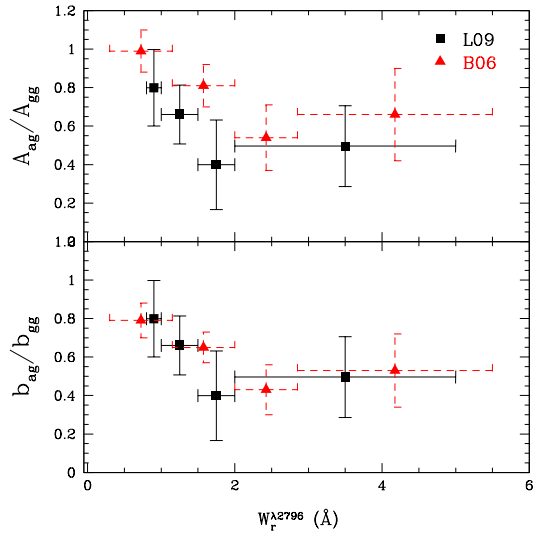}
\caption{Top: The clustering amplitude of  Mg II absorbers relative to the LRGs, $A_{ag}/\ A_{gg}$, measured for four sub-samples of  Mg II cut by rest equivalent width (solid, black).  For comparison, the results of B06 are overplotted (dashed, red). Bottom: The bias ratio $b_{ag}/\ b_{gg}$ of each equivalent width sample for this work, compared to the measurement of B06, which was corrected for stated overestimations resulting from photometric redshift uncertainties.}
\end{center}
\end{figure}
    
\section{Discussion}
\subsection{Comparison to Previous Results}

Generally, we confirm the B06 finding that $W_{r}\lesssim1.0$\AA $ $  Mg II absorbers cluster more strongly than those with $W_{r}\gtrsim1.5$\AA $ $, as shown in Figure 10 (see, Gauthier et al. 2009 for an additional confirmation of this result).  We do not fit our measurements on scales smaller than 1h$^{-1}$Mpc, but we find evidence for a sharp drop in the Mg II -- LRG cross-correlation for absorbers with $W_{r}<1.4$\AA $ $ on scales $\lesssim0.2$h$^{-1}$Mpc, which is not present in the $W_{r}\geq1.4$\AA $ $ measurement (see Figure 10).  The comoving distance at which this drop in amplitude is seen roughly coincides with the maximum extent of Mg II gas within the dark matter halo of a single galaxy ($\sim100$h$^{-1}$kpc; see, e.g., Chen \& Tinker 2008).  Since the weaker absorber sample exhibits stronger clustering as a whole, we can expect that the environments probed by these absorbers are more dense and thus more likely to obscure background quasars due to an increased number of foreground galaxies and extinction due to dust.   The effects of weak lensing must also be considered when interpreting these cross-correlation results on small scales, and we discuss this in detail in the following section. 

We have also obtained an estimate for the average  Mg II -- LRG bias ratio that is consistent, within the stated 1$\sigma$ errors, of the B06 measurements.  However there are a number of noteworthy differences in our samples and procedures that may explain some yet outstanding differences in our results, and we detail these below.

The amplitudes we measure for the  Mg II -- LRG cross-correlation and LRG auto-correlation are respectively $\sim13\%$ and $\sim45\%$  greater than those reported by B06 (see Table 6).  Differences in the measured cross- and auto-correlation amplitudes may be explained in part by a notable difference in the LRG redshift distribution of this paper and B06, which employed an alternate color selection algorithm (see Scranton et al. 2003) to that which we have adopted.  Differences in the LRG selection criteria are likely to produce a slightly different galaxy sample with a different average halo mass and clustering properties, which could explain discrepancies in the measured clustering amplitudes.  

It is worth noting that the LRG sample of B06 exhibits a similar redshift distribution to our sample for galaxies with z$<$0.65, but at high redshift the distributions noticeably diverge and a surplus of LRGs with photometric redshifts in the range $0.7 < z_{LRG} < 0.85$ appears in the B06 sample.  This excess can be explained by known stellar contamination of the high-redshift LRGs selected with the criteria employed by B06 (R. Scranton, private communication).   Any significant amount of stellar contamination will effectively reduce the amplitudes of both the  Mg II -- LRG  cross-correlation and the LRG auto-correlation.  However, Bouch{\'e} et al. (2004; 2005) have shown that stellar contamination of the LRG sample reduces these correlation amplitudes equally and ultimately cancels out in the measurement of the absorber -- galaxy amplitude ratio, $A_{ag}/A_{gg}$, since the same galaxy sample is applied to calculate both $A_{ag}$ and $A_{gg}$.  Therefore, we would not expect a contaminated galaxy sample to produce the lower values we obtain for $A_{ag}/A_{gg}$.  

\begin{figure}[t!]
\begin{center}
\plotone{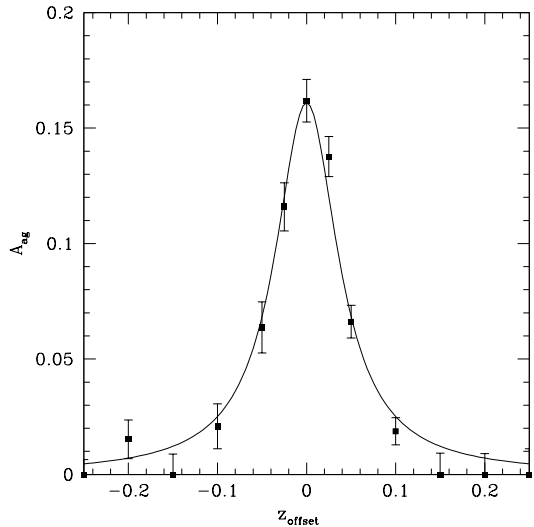}
\caption{The  Mg II -- LRG cross-correlation amplitude measured as a function of an imposed redshift offset.  A best-fit Lorentzian curve has been overplotted, with a FWHM=0.086.  From this test we extract a typical photometric redshift error of $\sigma_{z}=0.036$ for the LRG data used in this work.}
\end{center}
\end{figure}

Smaller photometric redshift errors propagate directly into higher measured correlation amplitudes, so it is reasonable to suggest that the improved techniques of Collister et al. (2007) that were used to estimate the LRG photometric redshifts in this work were the dominant factor contributing to both the higher correlation amplitudes we report and our lower value of $A_{ag}/A_{gg}$.  To explore this effect quantitatively, we have measured the  Mg II -- LRG cross-correlation amplitude as a function of an artificial redshift offset, $z_{offset}$, applied to the  Mg II absorbers.  The results are shown in Figure 12 with a best-fit Lorentzian curve with FWHM=0.086 overplotted.  Errors on the spectroscopic redshifts for QALs in the SDSS are of the order $\sigma_{z}=0.0001$, so the decreasing correlation amplitude with increasing $|z_{offset}|$ provides a measurement of the typical errors for the LRG photometric redshift estimates.  The FWHM of this distribution therefore corresponds to a typical photometric redshift uncertainty of $\sigma_{z}$ = FWHM/2.35 =  0.036.  This uncertainty agrees precisely with the variance of photometric redshift errors expected at the peak of the LRG redshift distribution (Collister et al. 2007).   It is furthermore worth noting that the effects of the absorber redshift offset presented in Figure 12 are symmetric with respect to the sign of the offset, indicating that there is no overall systematic bias in the LRG photometric redshifts to be over- or under-estimated.  The typical photometric redshift errors we estimate show a factor of 2 improvement compared with the errors reported by B06, which were determined using the same method.  


By comparing the observed absorber -- galaxy amplitude ratio, $A_{ag}/A_{gg}$ to expected results using mock galaxy catalogs B06 determined that their measurements of $A_{ag}/A_{gg}$ are overestimated by 25$\pm10$\% as a result of the LRG photometric redshift uncertainties.  More recently, Gauthier et al. (2009) undertook a similar analysis and found that with improved photometric redshifts similar to those used in this work, the overestimation should be at most a 10\% effect.  Thus, we do not apply a correction factor when converting from amplitude ratio to bias ratio and still find that the resulting bias ratio measurements of this work remain consistent with those of B06, as shown in Figure 11.


\begin{figure}[t!]
\begin{center}
\plotone{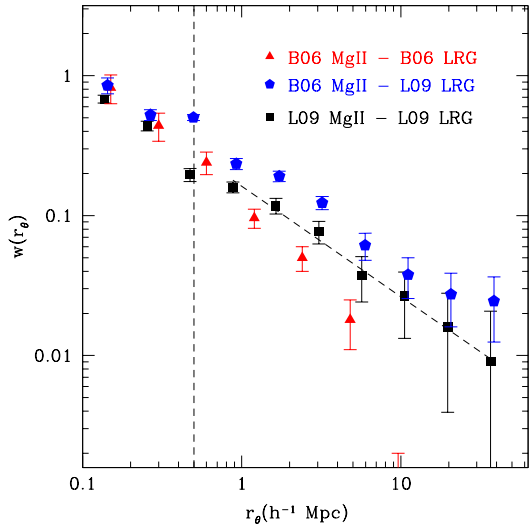}
\caption{A comparison of the W$_{r}\geq0.8$\AA $ $  Mg II -- LRG cross-correlation from Figure 7 (squares) with the measurements of B06 (triangles).  The results of cross-correlating the DR5 LRGs from this work with the  Mg II catalog of B06 are also shown (pentagons).}  
\end{center}
\end{figure}

In order to identify any differences in our measurements that have not resulted from the galaxy sample selection and photometric redshift estimation, we have cross-correlated the B06  Mg II catalog with the DR5 LRG sample used in this work.   In Figure 13 we compare our measurements of the  Mg II -- LRG cross-correlation to those of B06, and we overplot the results of cross-correlating the B06  Mg II sample with our DR5 LRGs.  As we might expect, the amplitude of the cross-correlation receives a strong boost from the improved photometric redshift estimates of the DR5 LRG sample.  However, this test also reveals that the B06  Mg II sample exhibits stronger clustering than the DR5 absorbers, even when cross-correlating with identical galaxies, indicating the existence of additional differences among the absorber samples.

As detailed in Section 2.3, in order to isolate any effects of weak lensing on the  Mg II -- LRG cross-correlation, we have required z$_{qso}$$>$0.8 in our Mg II sample.  We have also imposed a minimum velocity for the absorption systems in the quasar rest frame: $\beta_{min}=0.2$.  Doing so has removed any excess associated absorption lines we would expect to observe in the local environments of the quasars or as a result of high-velocity outflows from the central engine.  In comparison, the B06 sample includes low-redshift quasars (z$\sim$0.5), and 341 of their 1806 absorbers have $\beta<0.2$.

\begin{figure}[t!]
\begin{center}
\plotone{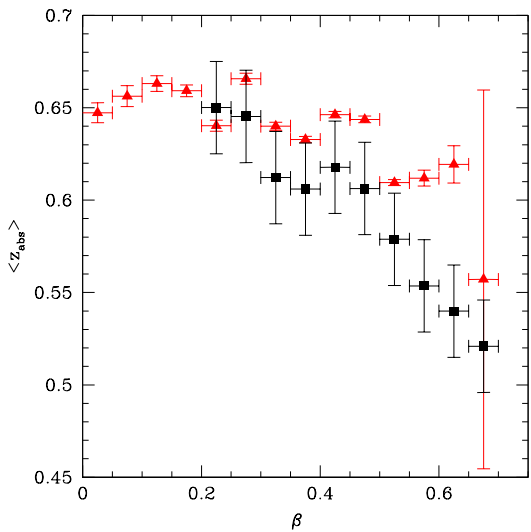}
\caption{A comparison of the mean redshifts and 1$\sigma$ errors for the DR5  Mg II sample (squares) and the B06  Mg II sample (triangles), plotted as a function of $\beta$.  The trend of decreasing mean redshifts with increasing $\beta$ results from the convolution of the redshift distributions of the absorbers and quasars.}
\end{center}
\end{figure}

The inclusion of  low-$\beta$ QALs can amplify the  Mg II -- LRG clustering signal in two ways.  For small velocity separations ($\beta<0.05$), measurements of the  Mg II -- LRG clustering may be biased by the underlying quasar -- LRG clustering.  However, this effect should only be significant on the smallest scales we examine.   Secondly, the redshift distribution of low-$\beta$ absorption systems traces that of the quasars themselves.  Since the number of quasars peaks at z$\gtrsim$1, the majority of low-$\beta$  Mg II systems will overlap in redshift-space with the highest-redshift LRGs.  To illustrate this, we cut the  Mg II absorber sample into velocity bins of width $\Delta\beta$=0.05 and computed the mean redshift for each sub-sample.  The results are shown graphically in Figure 14 for both our sample and that of B06.  As expected, lower velocity absorbers have significantly higher mean redshifts.  The decline in average redshift with increasing $\beta$ is stronger in our sample, due to our imposed restrictions of z$_{qso}>0.8$ and z$_{abs}\leq0.8$; but generally, we find that by requiring a large value of $\beta_{min}$ we have reduced the number of high-redshift absorbers for the redshift range and quasar sample examined in this work.

The method we employ to measure the  Mg II -- LRG cross-correlation is biased by the redshift distribution of the absorbers, as it measures the mean cross-correlation over the full redshift range of the sample.  If many more absorbers are included at one specific redshift, the mean cross-correlation of the sample will be weighted toward the typical LRG halo mass at that redshift.  We have shown that the choice of $\beta_{min}$ affects the redshift distribution of the absorbers, so it follows that this also affects the measurement of $\omega(r_{\theta})$.  While it has been shown that LRGs are generally non-evolving in stellar mass over the redshift range $0.4\lesssim z \lesssim0.8$ (Wake et al. 2006, 2008a; Brown et al. 2007; Cool et al. 2008; Brown et al. 2008), our ability to detect LRGs drops off sharply at high redshift.  Photometrically-selected LRGs at high redshifts must therefore be intrinsically more luminous for detection in a magnitude-limited survey such as the SDSS, and since luminosity and halo mass of LRGs have a well-established correlation (Zehavi et al. 2005), we should expect  that the LRGs on the high-redshift end of this sample inhabit the most massive dark matter haloes.  It follows that our estimates of the typical dark matter halo masses of the absorbers will be dependent on the convolution of the LRG and absorber redshift distributions.  However, provided the Mg II sub-samples we compare exhibit statistically similar redshift distributions, the absorber dark matter halo bias will be measured relative to the same typical LRG halo mass in each case. 

\begin{figure}[t!]
\begin{center}
\plotone{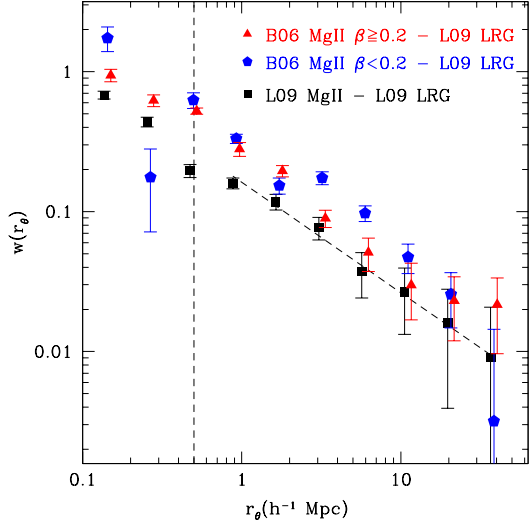}
\caption{A comparison of the DR5 W$_{r}\geq0.8$\AA $ $  Mg II -- LRG cross-correlation from Figure 7 (black)  with two B06  Mg II ---DR5 LRG cross-correlations, incorporating further cuts on the absorber velocity in the quasar rest-frame ($\beta$).  Measurements of the $\beta\leq0.2$  Mg II absorbers from the B06 catalog cross-correlated with DR5 LRGs are shown in blue.  B06  Mg II cut to the same velocity limit as absorbers in this work ($\beta>0.2$) are also cross-correlated with DR5 LRGs and shown in red.}
\end{center}
\end{figure}

To confirm that the choice of $\beta_{min}$ affects the measured clustering amplitude as expected, in Figure 15 we separate the 341 low-$\beta$ ($\beta<0.2$) systems in the B06 sample and cross-correlate these QALs with the DR5 LRGs used in this work.  The low-$\beta$ QALs appear to be more strongly clustered than both the high-$\beta$ systems and the average of the LRGs (see Figure 10).  In contrast, the clustering of the  1,125 high-$\beta$ QALs in the B06 sample agrees with the DR5 sample on scales $\sim$1$-$10 h$^{-1}$Mpc.  Our choice of a higher $\beta_{min}$ for this work has ensured that the absorber sample represents an unambiguously intervening population of  Mg II, and we find the this requirement to be the primary cause of the lower Mg II clustering amplitude we measure relative to the earlier results of B06.

 \begin{figure}[t!]
\begin{center}
\plotone{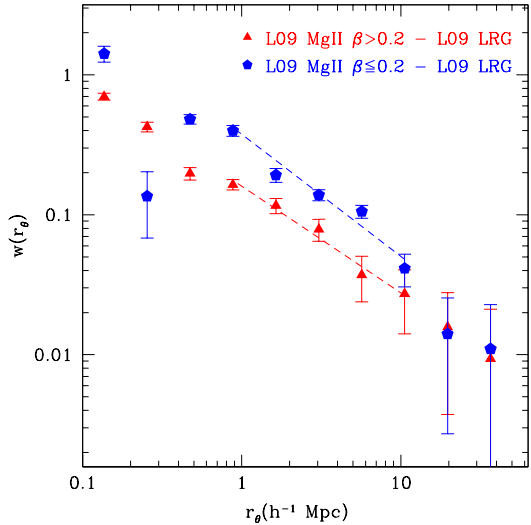}
\caption{The  Mg II --- LRG cross-correlation of two samples of DR5  Mg II QALs, cut by velocity in the quasar rest-frame ($\beta$).  Measurements using the $\beta\leq0.2$  Mg II absorbers are shown in blue.  Measurements using  Mg II QALs cut to the same velocity limit as the primary sample used in this work ($\beta>0.2$) are shown in red.  We find by imposing a minimum velocity of $\beta>0.2$ on our sample reduces the average clustering amplitude on scales 1-10 h$^{-1}$Mpc  by approximately a factor of 2.}
\end{center}
\end{figure}

In order to test this effect with greater statistical certainty, we repeat the aforementioned analysis using low- and high-$\beta$ samples from the data presented in Table 1.  Recall that in creating Table 1, we lifted the restrictions on the quasar redshift, thereby allowing a larger number of low-$\beta$ ($\beta<0.2$) QALs to be included in the analysis.  All other selection criteria described in Section 2.3 are retained.  The resulting samples contain 719  Mg II absorbers with $\beta<0.2$ and 2,750 with $\beta\geq0.2$.  We measure the cross-correlation of each sample with the LRGs, and present the results in Figure 16.  

The high-$\beta$ QALs exhibit power-law behavior over the range 0.5--40 h$^{-1}$Mpc and cross-correlation measurements that are completely consistent with our primary sample (as shown in Figure 7). In contrast, the low-$\beta$ systems produce a higher average clustering amplitude and exhibit a departure from power-law behavior on the smallest and largest scales.  The behavior is consistent with the cross-correlation measured for the B06  Mg II sample, shown in Figure 15.  These effects may be due to random scatter caused by the smaller samples of low-$\beta$ QALs available in each analysis.  However, because such similar behavior is found for both the B06 and DR5 data, it is likely that either (1) a more complex multiple halo structure is traced by these primarily high-redshift absorbers or (2) the low-$\beta$ systems, which are expected to contain significant contamination from quasar outflows, include redshifts that provide unreliable estimates of their true positions and therefore do not accurately trace large-scale structure.   Because the large and small-scale behavior is not well understood for the low-$\beta$ systems, we fit these measurements on scales $\sim$1-10 Mpc, where the cross-correlation behaves approximately as a power-law. 

 We find that the $\beta\geq0.2$ absorber sample returns a clustering amplitude of $A_{ag}=0.160\pm0.01$, while absorbers with $\beta<0.2$ have $A_{ag}=0.375\pm0.016$ -- approximately a factor of two higher than the intervening population.  The average redshift for the $\beta\geq0.2$ sample of 2,750 absorbers is z=0.596, which is nearly equivalent to the average redshift of the $\beta<0.2$ sample, z=0.616.  Thus we should not expect that the significantly higher clustering amplitudes of the $\beta<0.2$ sample simply results from redshift-dependent scaling of the average LRG dark matter halo mass in our magnitude-limited sample, as discussed previously.  

The simplest and most likely explanation for the higher clustering in the $\beta\leq0.2$ sample is that the population of associated Mg II absorbers is, on average, physically correlated with dark matter haloes of significantly higher mass compared to the population of unambiguously intervening absorbers.  We measure the amplitude ratio of the $\beta\leq0.2$ sample with respect to the LRGs to be $A_{ag}/A_{gg}=1.4\pm0.28$, implying that the population of associated absorbers typically inhabits dark matter haloes even more massive than those of average LRGs. 

If we again employ the same formalism used to approximate the dark matter halo masses of these absorbers, we measure a dark matter halo bias of 1.09$\pm$0.22 for the $\beta\geq0.2$ sample, corresponding to a dark matter halo mass of log$M_{h}(M_{\sun}h^{-1})=12.05\pm^{0.64}_{0.158}$; for the $\beta<0.2$ sample we calculate a bias of 2.55$\pm$0.51, indicating a dark matter halo mass of log$M_{h}(M_{\sun}h^{-1})=13.95\pm^{0.24}_{0.33}$.  Thus, the dark matter haloes hosting associated Mg II absorbers are typically $\sim$10-100 times more massive than those probed by intervening Mg II systems.  This mass estimate is not implausible, since in selecting a sample of associated absorbers we are likely already biasing our measurement to regions containing both a quasar and at least one neighboring galaxy.

A few notable recent results present additional evidence to support this finding of higher dark matter halo mass environments for associated Mg II absorbers.  Wild et al. (2008) report that $\sim$8\% of quasars in the DR5 exhibit associated Mg II absorption and that the velocity distribution of associated Mg II absorbers can be explained by physical clustering of satellite galaxies in the local quasar environment without invoking outflows.   Furthermore, an observed over-abundance of associated absorption in radio-loud quasars has been noted by Wild et al. (2008) and Richards et al. (1999).  Wake et al. (2008b) find that radio-loud LRGs are found in dark matter haloes that are approximately twice as massive than those of radio-quiet LRGs in the same redshift range of LRGs used in this analysis.  Similarly, radio-loud quasars have been shown to reside in haloes of mass log$M_{h}(M_{\sun}h^{-1})\sim13$, while radio-quiet quasars have typical halo masses of log$M_{h}(M_{\sun}h^{-1})\sim12.3$ (Shen et al. 2008).  Thus, it should not be a surprise to also find that associated Mg II absorption, which is often coincident with radio-loud quasars, is also associated with haloes of significantly higher mass.  Measurements of the large-scale clustering of quasars both with and without associated Mg II absorption should shed more light on this issue, which we intend to pursue in more detail in a follow-up paper.
 
 \begin{figure}[t!]
\begin{center}
\plotone{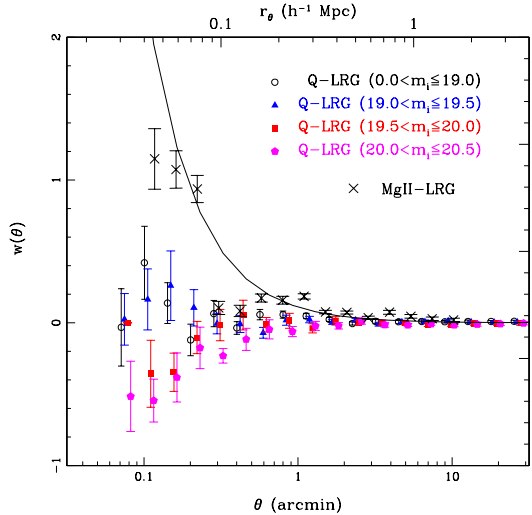}
\caption{The angular correlation of background DR5 quasars, cut into sub-samples by apparent $i$-band magnitude, with foreground LRGs.  We overplot the measured angular cross-correlation of the entire sample of 2,705  Mg II systems ($W_{r}>0.8$\AA) with LRGs and the corresponding minimum $\chi^{2}$ fit of these data to a power-law.  We find weak lensing can produce a bias in the  Mg II -- LRG projected cross-correlation on the 20-30\% level for angular scales $<2\arcmin$, corresponding to physical scales $<$0.5 h$^{-1}$Mpc for the average redshift of the absorber ($\overline{z}\sim0.6$).  The comoving distance scale, approximated at $\overline{z}$, is provided on the top axis.}
\end{center}
\end{figure}

 \subsection{Effects of Weak Lensing and Dust Extinction}
 
 The detection of QALs is dependent on the distribution and properties of background quasars.  As a result, selection biases inherent in the quasar sample will propagate into the QAL data as well.  Quasars are subject to weak lensing by foreground galaxies, which can amplify the apparent magnitudes we observe.  Furthermore, our ability to detect QALs is notably hindered by the poor signal-to-noise of the faintest SDSS quasars. Therefore, for completeness, we investigate how the quasar magnitude-dependence of the absorption line selection affects the measured absorber-LRG cross-correlation.  
 
We begin by calculating the angular cross-correlation of the background quasar sample and foreground LRGs.  For this analysis we include all quasars from the DR5 catalog \citep{S07} that have $z>0.8$ and meet the same seeing and reddening criteria applied to the absorber and LRG samples.  We make no added requirement on the presence or absence of absorption in these objects.  This sample of 46,747 quasars is cut into four sub-samples according to $i$-band magnitude: 13,171 with $0<m_{i}\leq19.0$, 16,481 with $19.0<m_{i}\leq19.5$, 8,862 with $19.5<m_{i}\leq20.0$, and 8,233 with $20.0<m_{i}\leq20.5$.  We then calculate the angular cross-correlation for each magnitude-limited sample of quasars with the DR5 LRG sample previously described.  Again, we use a variation on the estimator from Adelberger et al.  (2003):
\begin{equation}
\omega_{qg}(\theta)=\frac{QG}{QR} - 1.0
\end{equation}
where QG are the number of quasar-galaxy pairs within each angular scale, and QR is the number of quasar-random galaxy pairs, normalized by the ratio of the total number of real galaxies $N_{G}$ to the number of random galaxies $N_{R}$ contained in the jackknife region of each quasar.  This estimator does not provide the most rigorous measurement of the angular clustering for the SDSS spectroscopic quasar sample (see, e.g., Ross et al. 2007 for precision measurements using the Landy-Szalay (1993) estimator).  However, the symmetry with our earlier analysis makes this estimator suitable for evaluating the bias of our projected cross-correlation measurements.

Our results, shown in Figure 17, reveal that angular cross-correlation is strongly dependent on the $i$-band magnitude of the background quasar on scales smaller than 2\arcmin, which corresponds to $\sim0.5$ h$^{-1}$Mpc for the average redshift of  Mg II absorbers in this work, $z\sim0.6$.  Specifically, we find that quasars with $m_{i}\leq19.5$ are significantly correlated on these small scales with the LRGs, while those quasars with $m_{i}>19.5$ exhibit a significant anti-correlation with the same LRG population.  We emphasize that these quasars do not overlap in redshift space with the LRG population; therefore, this clustering signal must be due to cosmic magnification and not the actual physical clustering between quasars and LRGs.

\begin{figure}[t!]
\begin{center}
\plotone{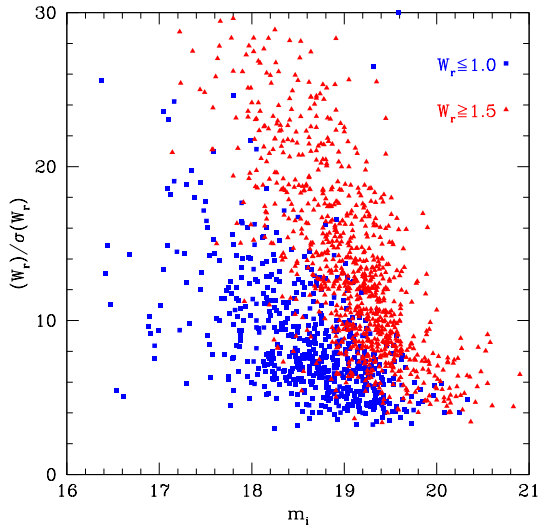}
\caption{The distribution of two equivalent width samples of Mg II, as functions of detection significance and m$_{i}$ of the background quasar.  In the brightest quasars, where the S/N should not bias the equivalent width distribution of detected lines, the majority of  Mg II lines are weak ($W_{r}^{\lambda2796}\leq1.0$\AA).  The fractional number of strong/weak detections increases dramatically for increasingly fainter quasars as a result of the SDSS S/N.}
\end{center}
\end{figure}

The detection of cosmic magnification was first unambiguously observed in the SDSS by Scranton et al. (2005).  This weak lensing effect causes quasars with massive foreground galaxies to be preferentially detected in a magnitude-limited sample (Narayan 1989).  While the flux of background quasars is increased by weak lensing effects to produce positive correlations with foreground galaxies on small scales, the faintest quasars will appear anti-correlated due to their preferential detection at high redshift, where their surface density is diluted by larger volumes of space.  We clearly observe this effect in our data.   

\begin{figure}[t!]
\begin{center}
\plotone{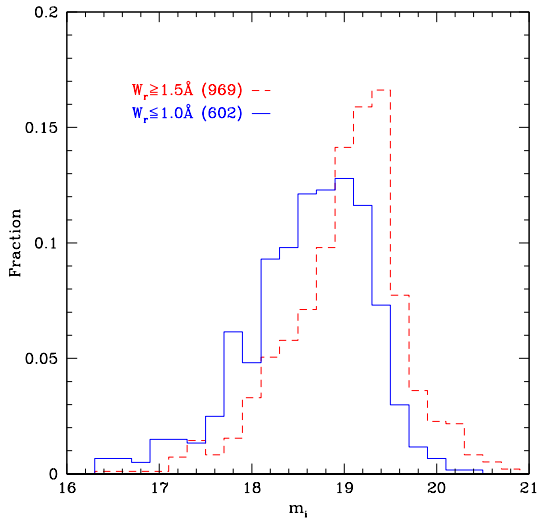}
\caption{The fractional $i$-band apparent magnitude distribution from two equivalent width samples of DR5 Mg II systems.  No requirement on $m_{i}$ has been imposed in this case, in contrast to Figure 9, which provides the distribution of strong and weak absorption lines observed in quasars with $m_{i}\leq19.5$.  By including quasars of all available magnitudes in this figure, we observe that the lower equivalent width lines are preferentially detected in the spectra of brighter quasars, due to their higher average signal-to-noise in the SDSS sample.}
\end{center}
\end{figure}

For comparison, in Figure 17 we also present the angular cross-correlation measured for the sample of 2,705  Mg II absorbers and 1,495,604 LRGs along with a minimum $\chi^{2}$ power-law fit to the results.  The absorbers and LRGs occupy the same redshift range in this case, so the amplitude of the cross-correlation is this time dominated by physical clustering.  However, since we find that weak lensing produces significant correlation in the quasar sample, this effect should also be contained within these  Mg II -- LRG angular cross-correlation measurements.  As stated in Section 2.3, the absorbers have all been extracted from quasars with $m_{i}<19.5$, so we expect that the absorber-LRG cross-correlation will be positively biased by weak lensing of the background quasars.  By comparing the amplitudes of the fits for the  Mg II -- LRG and quasar -- LRG angular cross-correlations, we find that the  Mg II -- LRG cross-correlation measurements for scales $<0.5$h$^{-1}$Mpc may be influenced by weak lensing on the level of $\sim20-30$\%.  As a result, we include only large scale ($r_{\theta}>0.5$h$^{-1}$Mpc) measurements in our fits to the  Mg II -- LRG projected cross-correlation in Section 4.

In Figure 18 we examine how the significance of line detections depends on both m$_{i}$ and W$_{r}$.  In this figure we include all of the most reliable z$<$0.8 Mg II systems in the Y09 catalog regardless of equivalent width of the W$_{r}^{\lambda2796}$ line (those with grades A or B), with no additional requirements on absorber velocity, quasar redshift, or apparent magnitude.  Of these systems, 602 have W$_{r}\leq1.0$\AA, and 969 have W$_{r}\geq1.5$\AA.  We find that in the spectra of the brightest quasars, the majority of the most reliable Mg II systems have W$_{r}^{\lambda2796}\leq1.0$\AA, but as one probes fainter and fainter quasars, this fraction turns over, until the overwhelming majority of detected  Mg II lines have W$_{r}^{\lambda2796}\geq1.5$\AA.  We should expect this effect since, given a constant integration time, brighter objects produce spectra with higher signal-to-noise, in which narrow lines are more easily detected.  This bias could compound the aforementioned lensing effects, by potentially producing an even stronger artificial clustering signal at small scales for the low equivalent width lines, which are more likely to be detected in the spectra of lensed background quasars.

For another view of this effect, we present in Figure 19 the fractional m$_{i}$ distribution for the same two equivalent width samples shown in Figure 18.  One clearly sees the peak of the QAL distribution m$_{i}$ shifting to lower values (brighter objects) for lower equivalent width samples.  In contrast, Figure 9 presents the distribution for each of the two equivalent width samples used in the correlation analysis.   As discussed in Section 4, the removal of lines with W$_{r}^{\lambda2796}<0.8$\AA $ $ or m$_{i}>19.5$ minimizes the difference in the mean m$_{i}$ of these samples.  As a result, we expect any weak lensing inherent in our measurements to have affected both of the equivalent width samples in this analysis similarly.

Although the gravitational lensing of background quasars by foreground galaxies with detected absorption has also been observed (e.g., Turnshek et al. 1997; Inada et al. 2007), the statistical significance of this effect is less certain due to the tendency of absorbers to be detected at large impact parameters from host galaxies, which minimizes the effects of weak lensing by absorbers in a large statistical sample.  Quasars with observed absorption have been shown to be statistically brighter (York et al. 1991; Vanden Berk et al. 1996; Richards et al. 1999; M{\'e}nard \& P{\'e}roux 2003; Murphy \& Liske 2004; Ellison et al. 2004;  Prochaska, Herbert-Fort \& Wolfe 2005), which has been interpreted as evidence of weak lensing.  However, as previously discussed, the increased signal-to-noise available for the spectra of brighter quasars substantially improves our ability to detect absorption, which complicates the certainty of interpreting these previous detections as weak lensing by absorbers.  In the largest statistical analysis to date, M{\'e}nard et al. (2008) were yet unable to detect any significant lensing of background quasars by foreground Mg II absorbers in the SDSS DR4.  This result implies that the lensing we observe in the quasar -- LRG cross-correlation shown in Figure 17 is dominated by the presence of foreground LRGs, and any lensing due to the incidence of Mg II in the quasar line of sight should be considered negligible.

It is well-established that quasar absorption systems contain dust, which reddens background quasars (e.g., York et al. 2006).  Since the SDSS quasar selection algorithm (Richards et al. 2002) targets quasars for spectroscopic followup based on specific color and magnitude requirements (Schneider et al. 2002), these reddening effects can cause quasars with strong absorbers to be selectively missed in the SDSS sample.  M{\'e}nard et al. (2008) quantified the effective reddening of background quasars by foreground Mg II absorbers as a function of equivalent width, finding that absorption strength correlates with the fraction of SDSS quasars missed due to extinction effects.  They find that for a sample of $W_{r}<1.0$\AA $ $, fewer than 1\% of quasars are missed, and fewer than 10\% of quasars are missed in a sample with $W_{r}<3.0$\AA $ $.  In this work, absorbers with $W_{r}<3.0$\AA $ $  comprise more than 95\% of our sample, and at the peak of our equivalent width distribution, $W_{r}\sim1.4$\AA $ $, coincides with a likelihood of missing 2\% of quasars, as determined by M{\'e}nard et al. (2008).  Therefore, we do not expect the effects of absorber-induced extinction to significantly affect our cross-correlation measurement.

 \subsection{Implications of $W_{r}^{\lambda2796}$--$Dependent$ $\partial N / \partial z$ Evolution}

As shown in Figures 4 and 5, the strongest Mg II absorbers in our sample exhibit redshift number density evolution, and the measured departure from a non-evolving $\partial N / \partial z$ becomes increasingly more significant as we examine samples with larger rest equivalent widths.  This result contrasts the findings of Nestor et al. (2005), which measured width-dependent evolution in  Mg II absorbers over the redshift range $0.5<z<2.0$ and found evidence of decreasing $\partial N / \partial z$, relative to a non-evolving behavior, with decreasing redshift for the strongest ($W_{r}^{\lambda2796}>2.0$\AA) lines.  The evolution we measure shows the opposite effect: namely, that the stronger lines in our sample exhibit more significant $increases$ in $\partial N / \partial z$, relative to a non-evolving behavior, with decreasing redshift.  

It is important to recognize that the total redshift range covered by our analysis is incorporated into a single redshift bin in the $\partial N / \partial z$ measurements of Nestor et al. (2005).  Thus, we have measured $\partial N / \partial z$  with much greater resolution across a smaller redshift range, and it is possible that our results are not in direct disagreement.  Furthermore, we have found that the lowest two redshift bins in our $\partial N / \partial z$ measurement suffer the most from contamination in the Mg II sample as a result of our inclusion of grade C systems, which are most often misidentified in the very blue end of the SDSS spectra.  However, removing these data points still does not significantly alter the results of the stated fits to non-evolution curves, so the effects of such contamination cannot alone explain the direction of departure from a non-evolving $\partial N / \partial z$ found among the strongest systems.   For the purposes of this work we are primarily concerned with how a width-dependent $\partial N / \partial z$ evolution may affect the absorber-galaxy cross-correlation.  Therefore, we reserve a more comprehensive analysis of the $\partial N / \partial z$, incorporating only the most certain detections in the SDSS DR5 QAL catalog, for future work.

In Section 5.1 we demonstrated that differences in absorber redshift distributions can propagate into the cross-correlation amplitude measurements; thus, a concern might be that any equivalent width-dependent $\partial N / \partial z$ evolution might similarly influence the relative bias measurements for the equivalent width samples presented in Figure 11.  We therefore make a series of cross-correlation measurements for the same four equivalent width-cut samples from Figures 5 and 11, after randomly removing absorbers until a flat redshift number distribution is achieved for each sample.  This process removes approximately 20\% of the absorbers in each sample, which increases the errors on the $\chi^{2}$ power law fits to the correlation functions by $\sim50\%$.  However, despite the loss in signal, we find no significant difference in the measurements of the relative dark matter bias for each equivalent width bin (shown for the complete samples in Figure 11).  

Since the equivalent width -- halo mass anti-correlation holds up in the absence of $\partial N / \partial z$ evolution for these absorbers, we may be seeing a bigger picture unfolding.  From the cross-correlation measurements presented in this work, it is clear that the 0.8\AA$\leq W_{r}^{\lambda2796}<1.0$\AA $ $ absorbers trace the highest mass dark matter haloes for absorbers in our sample and cluster similarly to the LRGs.  This finding fits well with the redshift number density evolution of these same absorbers, which exhibits no significant evolution over a range in redshift where the LRGs are similarly non-evolving.  In contrast, the largest equivalent width  Mg II absorbers in our sample ($W_{r}^{\lambda2796}>2$\AA) show both significant number density evolution for z$<$0.8 and lower relative clustering amplitudes, indicative of $\sim$25 times lower dark matter halo mass environments and consistent with field galaxies.  

These results are in agreement with the findings of Zibetti et al. (2007), which reported that the average color of light associated with weak ($W_{r}^{\lambda2796}\lesssim1.0$\AA)  Mg II systems is consistent with that of red passive galaxies, while the colors measured for strong ($W_{r}^{\lambda2796}\gtrsim1.5$\AA) systems match those observed in blue, star-forming galaxies.  If $W_{r}^{\lambda2796}\gtrsim2.0$\AA $ $ absorbers are typically found in dark matter halo environments of significantly lower mass compared to the $W_{r}^{\lambda2796}\lesssim1.4$\AA $ $ absorbers, as our cross-correlation results indicate, then the equivalent widths of these ultra-strong absorption systems are more likely the signature of outflows from regions of active star formation, rather than indicators of virialized gas in the most massive dark matter haloes.  

The global star formation rate is known to decline dramatically from z$\sim$1 to the present (Lilly et al. 1996; Madau et al. 1996), so the direction of the departure from non-evolution we observe with time in $\partial N / \partial z$ for the strongest Mg II systems is puzzling if these systems are indeed tracers of active star formation.  Models of cosmic downsizing (i.e., Cowie et al. 1996) propose that galaxies in lower mass haloes undergo active star formation at later epochs compared to their higher mass counterparts.  If the dark matter haloes of the strongest absorbers are indeed significantly less massive than the average LRG dark matter halo at the redshifts examined in this work, it is possible within the framework of cosmic downsizing that tracers of star formation in these lower mass galaxies could be more frequent at later times.  Measurements of the clustering amplitude for the strongest absorber sample over a range of redshifts would be needed to confirm this explanation.  For now we reserve a more detailed analysis of this issue for future work.

 \section{Conclusion}

We have presented the first absorber-galaxy cross-correlation analysis using the large QAL catalog of York et al. (2009, in prep.).  This has produced the current best measurement of the  Mg II -- LRG clustering amplitude and identified a number of critical biases to be considered in future absorber-galaxy correlation analyses.  Our primary findings and contributions to improving this analysis are itemized below.

\textbf{1. Equivalent-width dependence of the  Mg II -- LRG cross-correlation:}  Using the largest statistical sample to date, we confirm a previously reported weak anti-correlation of the equivalent width of Mg II absorption and dark matter halo mass, having measured the typical dark matter halo masses of Mg II absorbers to be: log$M_{h}(M_{\sun}h^{-1})=11.29\pm^{0.36}_{0.62}$ for a sample with W$_{r}$$\ge$1.4\AA $ $ and log$M_{h}(M_{\sun}h^{-1})=12.70\pm^{0.53}_{1.16}$  for absorbers with 0.8\AA$\leq W_{r}<$1.4\AA $ $. These observations imply that the weakest Mg II absorbers in our sample inhabit haloes with $\sim$25 times higher mass than the sample of strongest absorbers, in agreement with previous reported values of B06.  Measuring this effect in smaller equivalent width bins, we confirm a marginal anti-correlation of  Mg II equivalent width and clustering amplitude.  Although we have greatly refined the measurement of this effect, we do not recover a more significant relationship than that which was previously reported by B06.  

We do not fit our cross-correlation measurements on scales smaller than 1h$^{-1}$Mpc, but we find evidence for a significant decline in the cross-correlation of the weakest Mg II absorbers with LRGs on scales $\lesssim0.2$h$^{-1}$Mpc.  If these absorbers truly reside in higher mass haloes, this drop in correlation amplitude on the smallest scales could be explained by background quasars being reddened or obscured on impact parameters $\lesssim0.2$h$^{-1}$Mpc, as we would expect from the more significant presence of dust and other foreground galaxies in these environments.

\textbf{2.  Increased sample size:}  Compared to B06, we present a $\sim$50\% increase in the number of  Mg II systems (within a 37\% larger spectroscopic footprint, increasing from 4,188 deg$^{2}$ in the DR3 to 5,740 deg$^{2}$ in the DR5) and nearly a six-fold increase in the number of photometric LRGs (within a 51\% larger imaging footprint, increasing from 5,282 deg$^{2}$ in the DR3 to 8,000 deg$^{2}$ in the DR5).

\textbf{3.  Refinement of the LRG sample:}  The improved photometric redshift estimation algorithm of Collister et al. (2007), employed to produce the LRG catalog used in this work (Ross et al. 2008), provides approximately a factor of 2 reduction in the measured variance of the LRG photometric redshifts.  These improvements result in a 13\% higher Mg II - LRG cross-correlation amplitude, $A_{ag}$, and a 7\% lower overall measurement of the correlation amplitude ratio, $A_{ag}/A_{gg}$, compared with previous results (Bouch{\'e} et al. 2006).  

\textbf{4. Velocity distribution of  Mg II absorbers:} We apply velocity cuts justified by recent results (Richards et al. 1999; Nestor et al. 2008; Wild et al. 2008)  to remove associated absorption lines from the sample used in this clustering analysis and find that the higher correlation amplitude measurements previously reported for  Mg II may be explained by the inclusion of these associated absorbers.  We use the DR5 sample to measure the effect of imposing a conservative velocity cut and find that the cross-correlation amplitude of the associated absorber sample with $\beta\leq0.2$ is approximately twice the amplitude calculated for the unambiguously intervening ($\beta>0.2$) sample over this redshift range.  This difference corresponds to dark matter halo masses that are $\sim$100 times larger for the associated absorbers, relative to the intervening sample.  

\textbf{5. Implications of cosmic magnification:}  We determine that weak lensing of background quasars by foreground LRGs may significantly affect the  Mg II -- LRG clustering measurement on angular scales less than $\sim$2\arcmin, corresponding to a physical scale of $\sim0.5$h$^{-1}$Mpc, at the mean redshift of  Mg II QALs in our sample (z=0.61).  We also caution that weak lensing may preferentially affect the measurement of the equivalent width--halo mass correlation for the weakest lines of  Mg II in the SDSS, since the detection of weak lines strongly depends on signal-to-noise, and thereby also depends on the apparent magnitude of the quasar.  We have minimized this effect in our analysis by removing the faintest quasars (m$_{i}>$19.5) and imposing a minimum rest equivalent width of 0.8\AA.

\textbf{6. Evolution of $\partial N / \partial z$ as a function of $W_{r}^{\lambda2796}$:}  We find preliminary evidence to support stronger evolution in the redshift number density of strong ($>2.0$\AA)  Mg II absorption systems, relative to lower equivalent width samples ($\lesssim1.0$\AA) over the redshift range $0.4\lesssim z \lesssim0.8$.  This finding is consistent with the idea that these absorbers trace different galaxy populations.  Specifically, the weaker ($W_{r}^{\lambda2796}<1.4$\AA)  systems are associated with massive dark matter haloes with a non-evolving $\partial N / \partial z$, consistent with an LRG population; in contrast, the stronger ($W_{r}^{\lambda2796}\geq1.4$\AA) lines trace $\sim$25 times lower mass dark matter haloes and exhibit evolution in $\partial N / \partial z$ at a 2$\sigma$ significance level.  These findings may be consistent with models of cosmic downsizing in which lower-mass galaxies undergo active star-formation at progressively later times.  Measurements of the evolution of the clustering amplitudes of these strong absorbers should shed more light on this issue, and we reserve this analysis for future work.

\acknowledgments

BFL, RJB, AJR, and ADM acknowledge support from Microsoft Research, the University of Illinois, and NASA through grants NNG06GH156 and NB 2006-02049.  We thank David Wake, Peder Norberg, and Ryan Scranton for helpful discussions.  D.P.S. acknowledges support from NSF grant 06-07634.  Additionally, we thank Nicolas Bouch{\'e}, Michael Murphy, C{\'e}line P{\'e}roux, Istv{\'a}n Csabai, and Vivienne Wild for making their DR3 absorber catalog publicly available, and we thank Jean-Ren{\'e} Gauthier, Hsiao-Wen Chen and Jeremy Tinker for sharing their measurements of additional bias effects prior to publication.  The authors would also like to thank the anonymous referee for many helpful comments and suggestions.  BFL additionally thanks the University of Durham for their continued hospitality.

Funding for the SDSS and SDSS-II has been provided by the Alfred P. Sloan Foundation, the Participating Institutions, the National Science Foundation, the U.S. Department of Energy, the National Aeronautics and Space Administration, the Japanese Monbukagakusho, the Max Planck Society, and the Higher Education Funding Council for England. The SDSS Web Site is http://www.sdss.org/.

The SDSS is managed by the Astrophysical Research Consortium for the Participating Institutions. The Participating Institutions are the American Museum of Natural History, Astrophysical Institute Potsdam, University of Basel, Cambridge University, Case Western Reserve University, University of Chicago, Drexel University, Fermilab, the Institute for Advanced Study, the Japan Participation Group, Johns Hopkins University, the Joint Institute for Nuclear Astrophysics, the Kavli Institute for Particle Astrophysics and Cosmology, the Korean Scientist Group, the Chinese Academy of Sciences (LAMOST), Los Alamos National Laboratory, the Max-Planck-Institute for Astronomy (MPIA), the Max-Planck-Institute for Astrophysics (MPA), New Mexico State University, Ohio State University, University of Pittsburgh, University of Portsmouth, Princeton University, the United States Naval Observatory, and the University of Washington.

\clearpage

\begin{deluxetable}{cccccccccccccc}
\tablewidth{0pt}
\tablecaption{Excerpt from Electronic Catalog of Selected SDSS DR5 MgII Systems}
\tablecomments{The  3,467 MgII systems included in the electronic table are sorted by SDSS plate number and have been selected according to the following criteria: $W_{r}^{\lambda2796}\geq0.8$\AA, $m_{i}\leq19.5$, $z_{MgII}\leq0.8$, seeing$<$1.5$\arcsec$ and reddening$<$0.2 (using the masks of  Ross et al. 2006).  $DR$ represents the measured ratio of line strengths for each observed MgII doublet, $W_{r}^{\lambda2796}/W_{r}^{\lambda2803}$.}

\tablehead{
\colhead{RA} & 
\colhead{Dec} &
\colhead{$z_{MgII}$} & 
\colhead{$z_{QSO}$} & 
\colhead{$W_{r}^{\lambda2796}$(\AA)} & 
\colhead{$W_{r}/\sigma_{W_{r}}$} & 
\colhead{$DR$} & 
\colhead{$\sigma_{DR}$} & 
\colhead{$\beta$} & 
\colhead{$m_{i}$} & 
\colhead{plate} & 
\colhead{fiber} & 
\colhead{mjd}  &
\colhead{grade}
}

\tablecolumns{14}
\startdata
145.593801 & + 0.278311 & 0.5242 & 1.406 & 2.50 & 12.89 & 0.99 & 0.10 & 0.427 &  18.51  & 266 & 396 & 51630 & A \\
148.348608 & -0.634105 &  0.6379 & 1.382 & 1.66 & 18.44 & 1.30 & 0.12 & 0.358 & 18.56 & 267 & 169 & 51608 & A \\
150.073659 & +0.089913 & 0.6721 & 0.905 & 1.92 & 13.15 & 1.10 & 0.13 & 0.130 & 18.74 & 268 & 160 & 51633 & A \\
149.666926 & -0.155370 & 0.3981 & 1.585 & 2.71 & 7.45 & 1.51 & 0.26 & 0.547 & 18.69 & 268 & 193 & 51633 & B \\
149.083890 & -0.994077 & 0.6440 & 0.917 & 1.71 & 7.25 & 1.15 & 0.22 & 0.152 & 19.12 & 268 & 259 & 51633 & B \\
152.513102 & -0.225595 & 0.7052 & 0.732 & 1.26 & 13.40 & 1.06 & 0.11 & 0.016 & 18.71 & 270 & 191 & 51909 & A \\
152.155680 & -0.309766 & 0.6592 & 1.355 & 1.82 & 12.38 & 0.97 & 0.12 & 0.337  & 18.44 & 270 & 312 & 51909 & A \\
154.017760 & +0.203593 & 0.7063 & 1.483 & 0.81 & 6.75 & 1.72 & 0.48 & 0.358  & 18.46 & 271 & 399 & 51883 & B \\
154.033272 & +0.883398 & 0.5739 & 0.879 & 0.93 & 8.94 & 0.92 & 0.16 & 0.175 & 18.33  & 271 & 413 & 51883 & C \\
154.248255 & +0.378315 & 0.7103 & 2.090 & 0.80 & 10.53 & 1.19 & 0.20 & 0.531 & 18.62 & 271 & 432 & 51883 & C \\
156.709953 & +1.088432 &  0.4659 & 2.273 & 1.99 & 7.57 & 1.44 & 0.34 & 0.666 & 18.92  & 272 & 604 & 51941 & C \\
157.154241 & -1.007641 & 0.6322 & 1.532 & 1.61 & 14.38 & 1.14 & 0.12 & 0.413 & 17.91 & 273 & 286 & 51957 & A \\
157.154241 & -1.007641 & 0.7087 & 1.532 & 1.18 & 14.22 & 1.13 & 0.12 & 0.374 & 17.91 & 273 & 286 & 51957 & A \\
160.291050 & +0.181071 & 0.4766 & 2.259 & 2.11 & 14.86 & 1.25 & 0.15&  0.659 & 19.01  & 274 & 482 & 51913 & C \\
160.839918 & +0.722626 & 0.4631 & 0.624 & 2.48 & 14.42 & 1.05 & 0.10 & 0.104 & 18.96 & 274 & 584 & 51913 & B \\
. & . & . & . & . & . & . & . & . & . & . & . & . & . \\
. & . & . & . & . & . & . & . & . & . & . & . & . & . \\
\enddata
\end{deluxetable}
\begin{deluxetable}{ccccc}
\tablewidth{0pt}
\tablecaption{$\partial$N/$\partial$z Measurements for W$_{r}^{\lambda2796}$-limited Samples}
\tablecomments{$f_{N}$ represents the normalization factor determined from the minimum $\chi^{2}$ fit of each observed $\partial$N/$\partial$z to a non-evolving comoving density, $V_{c}(z)$.}

\tablehead{
\colhead{ z } &
\colhead{$\partial$N/$\partial$z} &
\colhead{$\partial$N/$\partial$z} &
\colhead{$\partial$N/$\partial$z} &
\colhead{$\partial$N/$\partial$z} \\
\colhead{($\pm$0.025)} &
\colhead{(W$>$0.8\AA)} &
\colhead{(W$>$1.0\AA)} &
\colhead{(W$>$1.5\AA)} &
\colhead{(W$>$2.0\AA)} 
}

\tablecolumns{5}
\startdata
 & & & & \\
0.3750 & 0.1259$\pm$0.0100 &  0.1031$\pm$0.0090 & 0.0535$\pm$0.0065 & 0.0260$\pm$0.0045 \\
0.4250 &  0.1939$\pm$0.0121 & 0.1561$\pm$0.0109 & 0.0905$\pm$0.0083 & 0.0483$\pm$0.0060 \\
0.4750 & 0.2260$\pm$0.0129 & 0.1810$\pm$0.0116 & 0.0938$\pm$0.0083 & 0.0473$\pm$0.0059 \\
0.5250 & 0.2106$\pm$0.0125 & 0.1583$\pm$0.0109 & 0.0829$\pm$0.0079 & 0.0448$\pm$0.0058 \\
0.5750 & 0.2455$\pm$0.0137 & 0.1931$\pm$0.0121 & 0.0935$\pm$0.0084 & 0.0471$\pm$0.0060 \\
0.6250 & 0.2550$\pm$0.0141 & 0.1948$\pm$0.0123 & 0.1048$\pm$0.0091 & 0.0540$\pm$0.0065 \\
0.6750 & 0.2816$\pm$0.0151 & 0.2273$\pm$0.0136 & 0.1331$\pm$0.0104 & 0.0698$\pm$0.0075 \\
0.7250 & 0.2957$\pm$0.0158 & 0.2169$\pm$0.0136 & 0.1093$\pm$0.0096 & 0.0610$\pm$0.0072 \\
0.7750 & 0.2907$\pm$0.0160 & 0.2236$\pm$0.0141 & 0.1096$\pm$0.0098 & 0.0557$\pm$0.0070 \\
& & & & \\
\hline
\hline
 & & & & \\
$f_{N}$ : & 1.289 & 0.999 & 0.522 & 0.272 \\

\enddata

\enddata

\end{deluxetable}

\begin{deluxetable}{ccccc}
\tablewidth{0pt}
\tablecaption{$\partial$N/$\partial$z Measurements for W$_{r}^{\lambda2796}$-limited Samples}

\tablehead{
\colhead{ z } &
\colhead{$\partial$N/$\partial$z} &
\colhead{$\partial$N/$\partial$z} &
\colhead{$\partial$N/$\partial$z} &
\colhead{$\partial$N/$\partial$z} \\
\colhead{($\pm$0.025)} &
\colhead{(0.8\AA$<$W$\leq$1.0\AA)} &
\colhead{(1.0\AA$<$W$\leq$1.5\AA)} &
\colhead{(1.5\AA$<$W$\leq$2.0\AA)} &
\colhead{(W$>$2.0\AA)} 
}

\tablecolumns{5}
\startdata
 & & & & \\
0.3750 & 0.0204$\pm$0.0040 & 0.0478$\pm$0.0061 & 0.0274$\pm$0.0046 & 0.0260$\pm$0.0045 \\
0.4250 &  0.0368$\pm$0.0053 & 0.0647$\pm$0.0070 & 0.0421$\pm$0.0056 & 0.0483$\pm$0.0060 \\
0.4750 &  0.0427$\pm$0.0056 & 0.0862$\pm$0.0080 & 0.0464$\pm$0.0059 & 0.0473$\pm$0.0059 \\
0.5250 &  0.0507$\pm$0.0061 & 0.0745$\pm$0.0075 & 0.0380$\pm$0.0053 & 0.0448$\pm$0.0058 \\
0.5750 &  0.0516$\pm$0.0063 & 0.0987$\pm$0.0087 & 0.0463$\pm$0.0059 & 0.0471$\pm$0.0060 \\
0.6250 &  0.0594$\pm$0.0068 & 0.0883$\pm$0.0083 & 0.0508$\pm$0.0063 & 0.0540$\pm$0.0065 \\
0.6750 &  0.0535$\pm$0.0066 & 0.0933$\pm$0.0087 & 0.0633$\pm$0.0072 & 0.0698$\pm$0.0075 \\
0.7250 &  0.0736$\pm$0.0079 & 0.1058$\pm$0.0095 & 0.0482$\pm$0.0064 & 0.0610$\pm$0.0072 \\
0.7750 &  0.0644$\pm$0.0075 & 0.1112$\pm$0.0099 & 0.0539$\pm$0.0069 & 0.0557$\pm$0.0070 \\
 & & & & \\
\hline
\hline
 & & & & \\
$f_{N}$ : & 0.288 & 0.467 & 0.249 & 0.272 \\

\enddata
\end{deluxetable}

\begin{deluxetable}{ccc}
\tablewidth{0pt}
\tablecaption{LRG Auto- and MgII-LRG Cross-Correlation Measurements}
\tablecomments{The MgII-LRG cross-correlation measurements, $\omega_{ag}$(r$_{\theta}$), in this table represent the population of absorbers with 0.8\AA$\leq W_{r}^{\lambda2796}<5.0$\AA. }
\tablehead{
\colhead{r$_{\theta}$ (h$^{-1}$Mpc)} &
\colhead{$\omega_{gg}$(r$_{\theta}$)} &
\colhead{$\omega_{ag}$(r$_{\theta}$)}
}
\tablecolumns{3}
\startdata
36.65 & 0.011$\pm$0.011 & 0.009$\pm$0.012 \\
19.68 & 0.018$\pm$0.012 & 0.016$\pm$0.012 \\
10.57 & 0.035$\pm$0.012 & 0.026$\pm$0.013 \\
5.68 & 0.063$\pm$0.012 & 0.038$\pm$0.013 \\
3.05 & 0.106$\pm$0.013 & 0.077$\pm$0.014 \\
1.64 & 0.169$\pm$ 0.014  & 0.118$\pm$0.015\\
0.88 & 0.268$\pm$0.015  & 0.160$\pm$0.014\\
0.47 & 0.482$\pm$0.018 & 0.196$\pm$0.021  \\
0.25 & 0.973$\pm$0.024 & 0.437$\pm$0.035 \\
0.14 & 1.970$\pm$0.038 & 0.680$\pm$0.043\\
\enddata

\end{deluxetable}
\begin{deluxetable}{ccc}
\tablewidth{0pt}
\tablecaption{MgII-LRG Cross-Correlation Measurements for W$_{r}^{\lambda2796}$-limited Samples}

\tablehead{
\colhead{r$_{\theta}$} &
\colhead{$\omega$(r$_{\theta}$)} &
\colhead{$\omega$(r$_{\theta}$)  } \\
\colhead{ (h$^{-1}$Mpc) } &
\colhead{ (0.8\AA$\leq W<1.4$\AA) } &
\colhead{ ($W\geq1.4$\AA) } 
}
\tablecolumns{3}
\startdata
 36.65 & 0.014$\pm$0.012 & 0.003$\pm$0.011 \\
19.68 & 0.022$\pm$0.012  & 0.009$\pm$0.012 \\
10.57 & 0.036$\pm$0.015 & 0.016$\pm$0.012 \\
5.68 & 0.048$\pm$0.015 & 0.026$\pm$0.012 \\
 3.05 & 0.085$\pm$0.015 & 0.068$\pm$0.014 \\
 1.64 & 0.151$\pm$0.021 & 0.080$\pm$0.011 \\
0.88 & 0.213$\pm$0.019 & 0.099$\pm$0.017 \\
0.47 & 0.196$\pm$0.021& 0.213$\pm$0.022  \\
 0.25 & 0.175$\pm$0.029 & 0.281$\pm$0.037 \\
 0.14 & 0.575$\pm$0.038 & 1.266$\pm$0.077\\

\enddata

\end{deluxetable}

\begin{deluxetable}{llcccc}
\tablewidth{0pt}
\tablecaption{Fit Parameters and Amplitude Ratio Measurements}
\tablecomments{Minimum $\chi^{2}$ fits of the MgII-LRG cross-correlation measurement have assumed a power-law model, $\widehat{\omega}=A\times(r_{\theta})^{\gamma}$, with 1$\sigma$ errors, fitted for scales $r_{\theta}>1.0$ h$^{-1}$Mpc.}
\tablehead{
\colhead{Figure } &
\colhead{Sample} &
\colhead{Size} &
\colhead{$A$} &
\colhead{$\gamma$} &
\colhead{$A_{ag}/A_{gg}$} 
}
\tablecolumns{6}
\startdata
7 & LRG & 1,495,605 & 0.268$\pm$0.052 & -0.886$\pm$0.033 &  $-$ \\
7 & MgII (0.8\AA$\leq W_{r}<5.0$\AA) & 2705 & 0.162$\pm$0.013 & -0.789$\pm$0.035  & 0.604$\pm$0.127 \\
 & & & & & \\
10 & MgII (0.8\AA$\leq W_{r}<1.4$\AA) & 1,444 & 0.192$\pm$0.029 & -0.749$\pm$0.161 & 0.716$\pm$0.176 \\
10 & MgII  (1.4\AA$\leq W_{r}<5.0$\AA) & 1,261 &  0.139$\pm$0.018 & -0.818$\pm$0.22 & 0.519$\pm$0.120 \\
 & & & & \\
11 &  MgII (0.8\AA$\leq W_{r}<1.0$\AA) & 590 & 0.214$\pm$0.034 & -0.602$\pm$0.134 & 0.799$\pm$0.193 \\
11 &  MgII (1.0\AA$\leq W_{r}<1.5$\AA) & 1,002 & 0.177$\pm$0.023 & -0.710$\pm$0.131 & 0.660$\pm$0.153 \\
11 &  MgII (1.5\AA$\leq W_{r}<2.0$\AA) & 541 & 0.107$\pm$0.059 & -0.918$\pm$0.411 & 0.399$\pm$0.233 \\
11 &  MgII (2.0\AA$\leq W_{r}<5.0$\AA) & 595 & 0.133$\pm$0.050 & -0.925$\pm$0.403 & 0.496$\pm$0.210 \\
\enddata

\end{deluxetable}

\clearpage


\begin{thebibliography}{}
\bibitem[Abazajian et al.(2003)]{2003AJ....126.2081A} Abazajian, K., et 
al.\ 2003, \aj, 126, 2081 
\bibitem[Abazajian et al.(2005)]{2005AJ....129.1755A} Abazajian, K., et al.\ 2005, \aj, 129, 1755 
\bibitem[Adelberger et al.(2003)]{2003ApJ...584...45A} Adelberger, K.~L., 
Steidel, C.~C., Shapley, A.~E., \& Pettini, M.\ 2003, \apj, 584, 45 
\bibitem[Adelman-McCarthy et al.(2007)]{AM07} Adelman-McCarthy, J.~K., et al.\ 2007, \apjs, 172, 634 
\bibitem[Bahcall(1968)]{Bahcall68} Bahcall, J.~N.\ 1968, \apj, 
153, 679 
\bibitem[Bahcall 
\& Spitzer(1969)]{1969ApJ...156L..63B} Bahcall, J.~N., \& Spitzer, L.~J.\ 1969, \apjl, 156, L63 
\bibitem[Bergeron(1986)]{1986A&A...155L...8B} Bergeron, J.\ 1986, \aap, 155, L8 
\bibitem[Bergeron \& Stasi{\'n}ska(1986)]{1986A&A...169....1B} Bergeron, J., \& Stasi{\'n}ska, G.\ 1986, \aap, 169, 1 
\bibitem[Bergeron et al.(1987)]{1987A&A...180....1B} Bergeron, J., Kunth, D., \&
D'Odorico, S.\ 1987, \aap, 180, 1
\bibitem[Berlind et al.(2003)]{2003ApJ...593....1B} Berlind, A.~A., et al.\
2003, \apj, 593, 1
\bibitem[Blake et al.(2008)]{2008MNRAS.385.1257B} Blake, C., Collister, A., 
\& Lahav, O.\ 2008, \mnras, 385, 1257 
\bibitem[Blanton et al.(2003)]{Blanton03} Blanton, M.~R., Lin,
H., Lupton, R.~H., Maley, F.~M., Young, N., Zehavi, I., \& Loveday, J.\
2003, \aj, 125, 2276  
\bibitem[Bouch{\'e} et al.(2004)]{2004MNRAS.354L..25B} Bouch{\'e}, N., 
Murphy, M.~T., \& P{\'e}roux, C.\ 2004, \mnras, 354, L25 
\bibitem[Bouch{\'e} et al.(2005)]{2005yCat..73549025B} Bouch{\'e}, N., Murphy, 
M.~T., \& Peroux, C.\ 2005, VizieR Online Data Catalog, 735, 49025 
\bibitem[Bouch{\'e} et al.(2006)]{B06} Bouch{\'e}, N., 
Murphy, M.~T., P{\'e}roux, C., Csabai, I., \& Wild, V.\ 2006, \mnras, 371, 
495 (B06)
\bibitem[Bouch{\'e} et al.(2007)]{2007yCat..73710495B} Bouch{\'e}, N., Murphy, 
M.~T., Peroux, C., Csabai, I., 
\& Wild, V.\ 2007, VizieR Online Data Catalog, 737, 10495
\bibitem[Brown et al.(2003)]{2003ApJ...597..225B} Brown, M.~J.~I., Dey, A.,
Jannuzi, B.~T., Lauer, T.~R., Tiede, G.~P.,
\& Mikles, V.~J.\ 2003, \apj, 597, 225
\bibitem[Brown et al.(2007)]{2007ApJ...654..858B} Brown, M.~J.~I., Dey, A., 
Jannuzi, B.~T., Brand, K., Benson, A.~J., Brodwin, M., Croton, D.~J., 
\& Eisenhardt, P.~R.\ 2007, \apj, 654, 858 
\bibitem[Brown et al.(2008)]{2008ApJ...682..937B} Brown, M.~J.~I., et al.\ 
2008, \apj, 682, 937 
\bibitem[Cannon et al.(2006)]{2006MNRAS.372..425C} Cannon, R., et al.\ 
2006, \mnras, 372, 425
\bibitem[Caulet(1989)]{1989ApJ...340...90C} Caulet, A.\ 1989, \apj, 340, 90 
\bibitem[Chelouche et al.(2008)]{2008ApJ...683...55C} Chelouche, D.,
M{\'e}nard, B., Bowen, D.~V., \& Gnat, O.\ 2008, \apj, 683, 55
\bibitem[Chen \& Tinker(2008)]{2008ApJ...687..745C} Chen, H.-W., \& Tinker, J.~L.\
2008, \apj, 687, 745
\bibitem[Churchill et al.(1999)]{1999ApJS..120...51C} Churchill, C.~W., 
Rigby, J.~R., Charlton, J.~C., \& Vogt, S.~S.\ 1999, \apjs, 120, 51
\bibitem[Churchill et al.(2000)]{2000ApJ...543..577C} Churchill, C.~W., 
Mellon, R.~R., Charlton, J.~C., Jannuzi, B.~T., Kirhakos, S., Steidel, 
C.~C., \& Schneider, D.~P.\ 2000, \apj, 543, 577
\bibitem[Churchill 
\& Vogt(2001)]{2001AJ....122..679C} Churchill, C.~W., \& Vogt, S.~S.\ 2001, \aj, 122, 679 
\bibitem[Churchill et al.(2005)]{2005ASPC..331..387C} Churchill, C., 
Steidel, C., \& Kacprzak, G.\ 2005, Extra-Planar Gas, 331, 387
\bibitem[Churchill et al.(2003)]{2003AJ....125...98C} Churchill, C.~W., 
Vogt, S.~S., \& Charlton, J.~C.\ 2003, \aj, 125, 98 
\bibitem[Collister et al.(2007)]{2007MNRAS.375...68C} Collister, A., et 
al.\ 2007, \mnras, 375, 68 
\bibitem[Cooke et al.(2006)]{2006ApJ...636L...9C} Cooke, J., Wolfe, A.~M., 
Gawiser, E., \& Prochaska, J.~X.\ 2006, \apjl, 636, L9 
\bibitem[Cool et al.(2008)]{2008ApJ...682..919C} Cool, R.~J., et al.\ 2008, 
\apj, 682, 919 
\bibitem[Cowie et al.(1996)]{1996AJ....112..839C} Cowie, L.~L., Songaila, 
A., Hu, E.~M., \& Cohen, J.~G.\ 1996, \aj, 112, 839 
\bibitem[Efron \& Gong (1983)]{}Efron, B., Gong, G., American Statistician, 37, 36
\bibitem[Eisenstein 
\& Hu(1998)]{1998ApJ...496..605E} Eisenstein, D.~J., \& Hu, W.\ 1998, \apj, 496, 605 
\bibitem[Eisenstein et al.(2001)]{eisenstein01} Eisenstein, D.~J.~et
  al.\ 2001, \aj, 122, 2267
  \bibitem[Eisenstein et al.(2005)]{2005ApJ...619..178E} Eisenstein, D.~J., 
Blanton, M., Zehavi, I., Bahcall, N., Brinkmann, J., Loveday, J., Meiksin, 
A., \& Schneider, D.\ 2005, \apj, 619, 178 
\bibitem[Eisenstein et al.(2005)]{2005ApJ...633..560E} Eisenstein, D.~J., 
et al.\ 2005, \apj, 633, 560 
\bibitem[Ellison et al.(2004)]{2004ApJ...615..118E} Ellison, S.~L., 
Churchill, C.~W., Rix, S.~A., \& Pettini, M.\ 2004, \apj, 615, 118
\bibitem[Elvis(2000)]{Elvis2000} Elvis, M.\ 2000, \apj, 545, 63 
\bibitem[Firth et al.(2003)]{2003MNRAS.339.1195F} Firth, A.~E., Lahav, O., 
\& Somerville, R.~S.\ 2003, \mnras, 339, 1195
\bibitem[Foltz et al.(1983)]{Foltz83} Foltz, C., Wilkes, B., 
Weymann, R., \& Turnshek, D.\ 1983, \pasp, 95, 341 
\bibitem[Fukugita et al.(1996)]{fukugita96} Fukugita, M.,
Ichikawa, T., Gunn, J.~E., Doi, M., Shimasaku, K., \& Schneider, D.~P.\
1996, \aj, 111, 1748
\bibitem[Ganguly et al.(2007)]{2007ApJ...665..990G} Ganguly, R., 
Brotherton, M.~S., Cales, S., Scoggins, B., Shang, Z., 
\& Vestergaard, M.\ 2007, \apj, 665, 990
\bibitem[Gauthier et al. (2009)] {} Gauthier, J.~R., Chen, H.~W., and Tinker, J.L., 2009, arXiv:0902.3237
\bibitem[Gibson et al.(2009)]{2009ApJ...692..758G} Gibson, R.~R., et al.\
2009, \apj, 692, 758
\bibitem[Gunn et al.(1998)]{gunn98} Gunn, J.~E.~et al.\ 1998,
  \aj, 116, 3040
\bibitem[Gunn et al.(2006)]{Gunn06} Gunn, J.~E., et al.\ 2006, 
\aj, 131, 2332
\bibitem[Hall et al.(2002)]{Hall02} Hall, P.~B., et al.\ 2002, 
\apjs, 141, 267 
\bibitem[Hamann(1997)]{1997ApJS..109..279H} Hamann, F.\ 1997, \apjs, 109, 
279 
\bibitem[Hogg et al.(2001)]{hogg01} Hogg, D.~W., Finkbeiner,
D.~P., Schlegel, D.~J., \& Gunn, J.~E.\ 2001, \aj, 122, 2129
\bibitem[Inada et al.(2007)]{2007AJ....133..206I} Inada, N., et al.\ 2007, 
\aj, 133, 206 
\bibitem[Ivezi{\'c} et al.(2004)]{ivezic042} Ivezi{\'c}, {\v Z}., 
et al.\ 2004, Astronomische Nachrichten, 325, 583 
\bibitem[Kacprzak et al.(2007)]{2007ApJ...662..909K} Kacprzak, G.~G., 
Churchill, C.~W., Steidel, C.~C., Murphy, M.~T., 
\& Evans, J.~L.\ 2007, \apj, 662, 909 
\bibitem[Landy 
\& Szalay(1993)]{1993ApJ...412...64L} Landy, S.~D., \& Szalay, A.~S.\ 1993, \apj, 412, 64 
\bibitem[Lanzetta et al.(1987)]{1987ApJ...322..739L} Lanzetta, K.~M., 
Wolfe, A.~M., \& Turnshek, D.~A.\ 1987, \apj, 322, 739 
\bibitem[Lanzetta 
\& Bowen(1990)]{1990ApJ...357..321L} Lanzetta, K.~M., \& Bowen, D.\ 1990, \apj, 357, 321 
\bibitem[Lilly et al.(1996)]{1996ApJ...460L...1L} Lilly, S.~J., Le Fevre, 
O., Hammer, F., \& Crampton, D.\ 1996, \apjl, 460, L1 
\bibitem[Lundgren et al.(2007)]{2007ApJ...656...73L} Lundgren, B.~F.,
Wilhite, B.~C., Brunner, R.~J., Hall, P.~B., Schneider, D.~P., York, D.~G.,
Vanden Berk, D.~E., \& Brinkmann, J.\ 2007, \apj, 656, 73
\bibitem[Lupton et al.(2001)]{lupton02} Lupton, R.~H., Ivezic,
Z., Gunn, J.~E., Knapp, G., Strauss, M.~A.,
\& Yasuda, N.\ 2002, \procspie, 4836, 350
  \bibitem[Madau et al.(1996)]{1996MNRAS.283.1388M} Madau, P., Ferguson, 
H.~C., Dickinson, M.~E., Giavalisco, M., Steidel, C.~C., 
\& Fruchter, A.\ 1996, \mnras, 283, 1388 
\bibitem[Magliocchetti \& Porciani(2003)]{2003MNRAS.346..186M} Magliocchetti, M.,
\& Porciani, C.\ 2003, \mnras, 346, 186
\bibitem[Maller \& Bullock(2004)]{2004MNRAS.355..694M} Maller, A.~H., \&
Bullock, J.~S.\ 2004, \mnras, 355, 694
\bibitem[Maller et al.(2005)]{2005ApJ...619..147M} Maller, A.~H., McIntosh,
D.~H., Katz, N., \& Weinberg, M.~D.\ 2005, \apj, 619, 147
\bibitem[M{\'e}nard 
\& P{\'e}roux(2003)]{2003A&A...410...33M} M{\'e}nard, B., \& P{\'e}roux, C.\ 2003, \aap, 410, 33 
\bibitem[M{\'e}nard et al.(2008)]{2008MNRAS.385.1053M} M{\'e}nard, B., 
Nestor, D., Turnshek, D., Quider, A., Richards, G., Chelouche, D., 
\& Rao, S.\ 2008, \mnras, 385, 1053 
\bibitem[Mo \& White(1996)]{1996MNRAS.282..347M} Mo, H.~J., \& White, S.~D.~M.\ 1996, \mnras, 282, 347 
\bibitem[Mo \& White(2002)]{2002MNRAS.336..112M} Mo, H.~J., \& White,
S.~D.~M.\ 2002, \mnras, 336, 112
\bibitem[Mo \& Miralda-Escude(1996)]{1996ApJ...469..589M} Mo, H.~J., \& Miralda-
Escude, J.\ 1996, \apj, 469, 589
\bibitem[Morton(2003)]{2003ApJS..149..205M} Morton, D.~C.\ 2003, \apjs, 
149, 205 
\bibitem[Murphy 
\& Liske(2004)]{2004MNRAS.354L..31M} Murphy, M.~T., \& Liske, J.\ 2004, \mnras, 354, L31 
\bibitem[Murray \& Chiang(1995)]{MC95} Murray, N., \& Chiang, J.\ 1995, \apjl, 454, L105
\bibitem[Myers et al.(2005)]{2005MNRAS.359..741M} Myers, A.~D., Outram, 
P.~J., Shanks, T., Boyle, B.~J., Croom, S.~M., Loaring, N.~S., Miller, L., 
\& Smith, R.~J.\ 2005, \mnras, 359, 741 
\bibitem[Narayan(1989)]{1989ApJ...339L..53N} Narayan, R.\ 1989, \apjl, 339, 
L53 
\bibitem[Navarro et al.(1996)]{1996ApJ...462..563N} Navarro, J.~F., Frenk, 
C.~S., \& White, S.~D.~M.\ 1996, \apj, 462, 563 
\bibitem[Nestor et al.(2005)]{2005ApJ...628..637N} Nestor, D.~B., Turnshek, 
D.~A., \& Rao, S.~M.\ 2005, \apj, 628, 637 
\bibitem[Nestor et al.(2007)]{2007ApJ...658..185N} Nestor, D.~B., Turnshek, 
D.~A., Rao, S.~M., \& Quider, A.~M.\ 2007, \apj, 658, 185 
\bibitem[Nestor et al.(2008)]{2008MNRAS.386.2055N} Nestor, D., Hamann, F., 
\& Hidalgo, P.~R.\ 2008, \mnras, 386, 2055 
\bibitem[Padmanabhan et al.(2005)]{2005MNRAS.359..237P} Padmanabhan, N., et 
al.\ 2005, \mnras, 359, 237 
\bibitem[Peebles(1974)]{1974A&A....32..197P} Peebles, P.~J.~E.\ 1974, \aap, 32, 197 
\bibitem[Petitjean \& Bergeron(1990)]{1990A&A...231..309P} Petitjean, P., \& Bergeron, J.\ 1990, \aap, 231, 309 
\bibitem[Pier et al.(2003)]{pier03} Pier, J.~R., et~al.\ 2003, \aj, 125, 1559
\bibitem[Prochaska 
\& Herbert-Fort(2004)]{2004PASP..116..622P} Prochaska, J.~X., \& Herbert-Fort, S.\ 2004, \pasp, 116, 622 
\bibitem[Prochaska et al.(2005)]{2005ApJ...635..123P} Prochaska, J.~X., 
Herbert-Fort, S., \& Wolfe, A.~M.\ 2005, \apj, 635, 123 
\bibitem[Prochter et al.(2006)]{2006ApJ...639..766P} Prochter, G.~E., 
Prochaska, J.~X., \& Burles, S.~M.\ 2006, \apj, 639, 766
\bibitem[Proga et al.(2000)]{Proga2000} Proga, D., Stone, J.~M., 
\& Kallman, T.~R.\ 2000, \apj, 543, 686
\bibitem[Reichard et al.(2003)]{Reichard03} Reichard, T.~A., et 
al.\ 2003, \aj, 126, 2594
\bibitem[Richards et al.(1999)]{1999ApJ...513..576R} Richards, G.~T., York, 
D.~G., Yanny, B., Kollgaard, R.~I., Laurent-Muehleisen, S.~A., 
\& vanden Berk, D.~E.\ 1999, \apj, 513, 576 
\bibitem[Richards et al.(2002b)]{richards02b} Richards, G.~T.~et
al.\ 2002, \aj, 123, 2945
\bibitem[Ross et al.(2006)]{2006ApJ...649...48R} Ross, A.~J., Brunner, 
R.~J., \& Myers, A.~D.\ 2006, \apj, 649, 48 
\bibitem[Ross et al.(2008)]{2008ApJ...682..737R} Ross, A.~J., Brunner, 
R.~J., \& Myers, A.~D.\ 2008, \apj, 682, 737 
\bibitem[Ross et al.(2007)]{2007MNRAS.381..573R} Ross, N.~P., et al.\ 2007, 
\mnras, 381, 573 
\bibitem[Ryan-Weber(2006)]{2006MNRAS.367.1251R} Ryan-Weber, E.~V.\ 2006, 
\mnras, 367, 1251
\bibitem[Sargent et al.(1988)]{1988ApJ...334...22S} Sargent, W.~L.~W., 
Steidel, C.~C., \& Boksenberg, A.\ 1988, \apj, 334, 22 
\bibitem[Schneider et al.(2002)]{2002AJ....123..567S} Schneider, D.~P., et
al.\ 2002, \aj, 123, 567
\bibitem[Schneider et al.(2007)]{S07} Schneider, D.~P., et 
al.\ 2007, \aj, 134, 102 
\bibitem[Scranton et al.(2003)]{S03} Scranton, R., et al.\ 
2003, ArXiv Astrophysics e-prints, arXiv:astro-ph/0307335 
\bibitem[Scranton et al.(2005)]{2005ApJ...633..589S} Scranton, R., et al.\ 
2005, \apj, 633, 589 
\bibitem[Shen et al.(2008)]{2008arXiv0810.4144S} Shen, Y., et al.\ 2008, 
arXiv:0810.4144 
\bibitem[Sheth et al.(2001)]{2001MNRAS.323....1S} Sheth, R.~K., Mo, H.~J., 
\& Tormen, G.\ 2001, \mnras, 323, 1 
\bibitem[Smith et al.(2002)]{smith02} Smith, J.~A.~et al.\
  2002, \aj, 123, 2121
\bibitem[Steidel 
\& Sargent(1992)]{1992ApJS...80....1S} Steidel, C.~C., \& Sargent, W.~L.~W.\ 1992, \apjs, 80, 1 
\bibitem[Steidel et al.(1994)]{1994ApJ...437L..75S} Steidel, C.~C., 
Dickinson, M., \& Persson, S.~E.\ 1994, \apjl, 437, L75 
\bibitem[Steidel et al.(1995)]{1995AJ....110.2519S} Steidel, C.~C., 
Pettini, M., \& Hamilton, D.\ 1995, \aj, 110, 2519 
\bibitem[Stoughton et al.(2002)]{Stoughton02} Stoughton, C., et 
al.\ 2002, \aj, 123, 485 
\bibitem[Strauss et al.(2002)]{strauss02} Strauss, M.~A.~et al.\
  2002, \aj, 124, 1810
  \bibitem[Tinker 
\& Chen(2008)]{2008ApJ...679.1218T} Tinker, J.~L., \& Chen, H.-W.\ 2008, \apj, 679, 1218 
\bibitem[Totsuji 
\& Kihara(1969)]{1969PASJ...21..221T} Totsuji, H., \& Kihara, T.\ 1969, \pasj, 21, 221 
\bibitem[Trump et al.(2006)]{2006ApJS..165....1T} Trump, J.~R., et al.\
2006, \apjs, 165, 1
\bibitem[Tucker et al.(2006)]{tucker06} Tucker, D.~L., et al.\ 
2006, Astronomische Nachrichten, 327, 821 
\bibitem[Turnshek et al.(1997)]{1997ApJ...485..100T} Turnshek, D.~A., 
Lupie, O.~L., Rao, S.~M., Espey, B.~R., 
\& Sirola, C.~J.\ 1997, \apj, 485, 100 
\bibitem[Tytler et al.(1987)]{1987ApJS...64..667T} Tytler, D., Boksenberg, 
A., Sargent, W.~L.~W., Young, P., \& Kunth, D.\ 1987, \apjs, 64, 667  
\bibitem[Vanden Berk et al.(1996)]{1996ApJ...469...78V} Vanden Berk, D.~E., 
Quashnock, J.~M., York, D.~G., \& Yanny, B.\ 1996, \apj, 469, 78 
\bibitem[Vanden Berk et al.(2008)]{2008ApJ...679..239V} Vanden Berk, D., et 
al.\ 2008, \apj, 679, 239 
\bibitem[Wake et al.(2006)]{2006MNRAS.372..537W} Wake, D.~A., et al.\ 2006, 
\mnras, 372, 537 
\bibitem[Wake et al.(2008a)]{2008MNRAS.387.1045W} Wake, D.~A., et al.\ 2008, 
\mnras, 387, 1045 
\bibitem[Wake et al.(2008b)]{2008MNRAS.tmp.1322W} Wake, D.~A., Croom, S.~M., 
Sadler, E.~M., \& Johnston, H.~M.\ 2008, \mnras, 1322
\bibitem[Weymann et al.(1979)]{1979ApJ...234...33W} Weymann, R.~J., 
Williams, R.~E., Peterson, B.~M., \& Turnshek, D.~A.\ 1979, \apj, 234, 33
\bibitem[Weymann et al.(1991)]{Weymann91} Weymann, R.~J., Morris, 
S.~L., Foltz, C.~B., \& Hewett, P.~C.\ 1991, \apj, 373, 23 
\bibitem[Wild et al.(2008)]{2008MNRAS.388..227W} Wild, V., et al.\ 2008, 
\mnras, 388, 227 
\bibitem[Wilman et al.(2007)]{2007MNRAS.375..735W} Wilman, R.~J., Morris, 
S.~L., Jannuzi, B.~T., Dav{\'e}, R., 
\& Shone, A.~M.\ 2007, \mnras, 375, 735
\bibitem[York et al.(1991)]{1991MNRAS.250...24Y} York, D.~G., Yanny, B., 
Crotts, A., Carilli, C., Garrison, E., \& Matheson, L.\ 1991, \mnras, 250, 24 
\bibitem[York et al.(2000)]{York00} York, D.~G., et al.\ 2000, 
\aj, 120, 1579 
\bibitem[York et al.(2009)]{York09} York, D.~G., et al. \ 2009, in preparation. (Y09)
\bibitem[Yuan \& Wills(2003)]{YW03} Yuan, M.~J., \& Wills, 
B.~J.\ 2003, \apjl, 593, L11
\bibitem[Zehavi et al.(2002)]{2002ApJ...571..172Z} Zehavi, I., et al.\ 
2002, \apj, 571, 172 
\bibitem[Zehavi et al.(2004)]{2004ApJ...608...16Z} Zehavi, I., et al.\ 
2004, \apj, 608, 16 
\bibitem[Zehavi et al.(2005)]{2005ApJ...621...22Z} Zehavi, I., et al.\ 
2005, \apj, 621, 22 
\bibitem[Zheng et al.(2008)]{2008arXiv0809.1868Z} Zheng, Z., Zehavi, I.,
Eisenstein, D.~J., Weinberg, D.~H., \& Jing, Y.\ 2008, arXiv:0809.1868
\bibitem[Zheng(2004)]{2004ApJ...610...61Z} Zheng, Z.\ 2004, \apj, 610, 61
\bibitem[Zibetti et al.(2005)]{2005ApJ...631L.105Z} Zibetti, S., 
M{\'e}nard, B., Nestor, D., \& Turnshek, D.\ 2005, \apjl, 631, L105 
\bibitem[Zibetti et al.(2007)]{2007ApJ...658..161Z} Zibetti, S., 
M{\'e}nard, B., Nestor, D.~B., Quider, A.~M., Rao, S.~M., 
\& Turnshek, D.~A.\ 2007, \apj, 658, 161 
\end{thebibliography}
\end{document}